\documentclass{jfm}
\usepackage{color}%highlight changes in revision

\usepackage{graphicx}
\usepackage{natbib}
\usepackage{epstopdf}
\usepackage{amsmath, amssymb}
\usepackage{xspace}%to get spacing right in new commands like e.g.
\usepackage{tikz}
\usepackage{pgflibraryshapes}
\usetikzlibrary{positioning}%fancy positioning
\usetikzlibrary{backgrounds}%box around tikz figure
\usetikzlibrary{calc}%coordinate calculations
\usepackage{titleref}\usepackage{byname}

\def \omegaNL {\mbox{$\omega_{\mbox{\tiny{NL}}}$}}
\def \ECHM {\mbox{$E_{\mbox{\tiny{CHM}}}$}}
\def \ZCHM {\mbox{$Z_{\mbox{\tiny{CHM}}}$}}
\def \rhovec { \mbox{\boldmath $\rho$} }

\def \ellvec { \mbox{\boldmath $\ell$} }
\def \e { \mbox{$\mathrm{e}$} }
\def \varepsilonE {\varepsilon_{\mbox{\tiny{$E$}}}}
\def \varepsilonW {\varepsilon_{\mbox{\tiny{$W$}}}}
\def \Sthree { \mbox{$\boldsymbol{S}_3^{\mbox{\tiny{$(W)$}}}$}}
\def \SthreePl { \mbox{$S_{3}^{\mbox{\tiny{$(W)$}}}$}}

\def \SthreeE { \mbox{$\boldsymbol{S}_3^{\mbox{\tiny{$(E)$}}}$}}
\def \SthreeEPl { \mbox{$S_3^{\mbox{\tiny{$(E)$}}}$}}
\def \Stwo { \mbox{$S_2$}}
\def \StwoE { \mbox{$S_{2E}$}}
\def \Source {\mbox{$\mathcal{F}_{\mbox{\tiny{}}}$}}
\def \SourceCHM {\mbox{$\mathcal{F}_{\mbox{\tiny{CHM}}}$}}
\def \SourceE {\mbox{$\mathcal{F}_{\mbox{\tiny{$E$}}}$}}
\def \ellF {\mbox{$\ell_i$}}
\def \hE {\mbox{$h_{\mbox{\tiny{$E$}}}$}}

\def \vth {v_{\mbox{\scriptsize{th}}}}
\def \i {\dot{\imath}}

\def \phi {\mbox{$\varphi$}}

\newcommand{\gyroavg}[1]{\left\langle#1\right\rangle_{\bf R}}
\newcommand{\gyroavgP}[1]{\left\langle#1\right\rangle_{{\bf R}^{\prime}}}
\newcommand{\angleavg}[1]{\left\langle#1\right\rangle_{\bf r}}
\newcommand{\angleavgP}[1]{\left\langle#1\right\rangle_{{\bf r}^{\prime}}}

\newcommand{\ensbl}[1]{\overline{#1}}
\newcommand{\CollisionOp}[1]{\gyroavg{C[#1]}}
\newcommand{\CollisionOpP}[1]{\gyroavgP{C^{\prime}[#1]}}
\newcommand{\CollisionOpZ}[1]{C[#1]}
\newcommand{\field}[1]{\mathbb{#1}}

\ifCUPmtlplainloaded \else
  \checkfont{eurm10}
  \iffontfound
    \IfFileExists{upmath.sty}
      {\typeout{^^JFound AMS Euler Roman fonts on the system,
                   using the 'upmath' package.^^J}%
       \usepackage{upmath}}
      {\typeout{^^JFound AMS Euler Roman fonts on the system, but you
                   dont seem to have the}%
       \typeout{'upmath' package installed. JFM.cls can take advantage
                 of these fonts,^^Jif you use 'upmath' package.^^J}%
      }
  \else
  \fi
\fi

\ifCUPmtlplainloaded \else
  \checkfont{msam10}
  \iffontfound
    \IfFileExists{amssymb.sty}
      {\typeout{^^JFound AMS Symbol fonts on the system, using the
                'amssymb' package.^^J}%
       \usepackage{amssymb}%

      }{}
  \fi
\fi

\ifCUPmtlplainloaded \else
  \IfFileExists{amsbsy.sty}
    {\typeout{^^JFound the 'amsbsy' package on the system, using it.^^J}%
     \usepackage{amsbsy}}
    {\providecommand\boldsymbol[1]{\mbox{\boldmath $##1$}}}
\fi

\providecommand\bnabla{\boldsymbol{\nabla}}

\newsavebox{\astrutbox}
\sbox{\astrutbox}{\rule[-5pt]{0pt}{20pt}}

\newcommand\etc{\textit{etc.}\xspace}
\newcommand\eg{\textit{e.g.}\xspace}
\newcommand\ie{\textit{i.e.}\xspace}

\title[Two-dimensional gyrokinetic turbulence]{Two-dimensional gyrokinetic turbulence}

\author[G. G. Plunk, S. C. Cowley, A. A. Schekochihin and T. Tatsuno]%
{G.\ns G.\ns P\ls L\ls U\ls N\ls K$^{1,2, 3}$ \thanks{Email: {\tt gplunk@umd.edu}},\ns
S.\ns C.\ns C\ls O\ls W\ls L\ls E\ls Y$^{4,5}$,
A.\ns A.\ns S\ls C\ls H\ls E\ls K\ls O\ls C\ls H\ls I\ls H\ls I\ls N$^{5, 6, 7}$ \and T.\ns T\ls A\ls T\ls S\ls U\ls N\ls O$^1$}
\affiliation{$^1$Department of Physics and IREAP, University of Maryland, College Park, MD 20742, USA\\[\affilskip]
$^2$Department of Physics, University of California, Los Angeles, CA 90095, USA\\[\affilskip]
$^3$Wolfgang Pauli Institute, University of Vienna, A-1090 Vienna, Austria\\[\affilskip]
$^4$EURATOM/CCFE Association, Culham Science Center, Abington OX1 3DB, UK\\[\affilskip]
$^5$Plasma Physics, Blackett Laboratory, Imperial College,  London SW7 2AZ, UK\\[\affilskip]
$^6$Rudolf Peierls Centre for Theoretical Physics, University of Oxford, Oxford OX1 3NP, UK\\[\affilskip]
$^7$Institut Henri Poincar\'e, Universit\'e Pierre et Marie Curie, 75231 Paris Cedex 5, France}

\pubyear{2010}
\volume{}
\pagerange{}
\begin{document}

\maketitle

%Force footnotes to be numbered.  The default is an arbitrary sequence of symbols.
\renewcommand{\thefootnote}{\textsuperscript{\arabic{footnote}}}

\begin{abstract}
Two-dimensional gyrokinetics is a simple paradigm for the study of kinetic magnetised plasma turbulence.  In this paper, we present a comprehensive theoretical framework for this turbulence.  We study both the inverse and direct cascades (the `dual cascade'), driven by a homogeneous and isotropic random forcing.  The key characteristic length of gyrokinetics, the Larmor radius, divides scales into two physically distinct ranges.  For scales larger than the Larmor radius, we derive the familiar Charney--Hasegawa--Mima (CHM) equation from the gyrokinetic system, and explain its relationship to gyrokinetics.  At scales smaller than the Larmor radius, a dual cascade occurs in phase space (two dimensions in position space plus one dimension in velocity space) via a nonlinear phase-mixing process.  We show that at these sub-Larmor scales, the turbulence is self-similar and exhibits power law spectra in position and velocity space.  We propose a Hankel-transform formalism to characterise velocity-space spectra.  We derive the exact relations for third-order structure functions, analogous to Kolmogorov's four-fifths and Yaglom's four-thirds laws and valid at both long and short wavelengths.  We show how the general gyrokinetic invariants are related to the particular invariants that control the dual cascade in the long- and short-wavelength limits.  We describe the full range of cascades from the fluid to the fully kinetic range.
\end{abstract}

\section{Introduction}

\subsection{Background}

A fluid is conventionally described by macroscopic state variables such as bulk flow velocity, density, pressure, \etc, which vary over three-dimensional space.  This description is appropriate when collisions between the constituent particles establish local thermodynamic equilibrium (Maxwellian velocity distribution) more rapidly than any dynamical processes can disturb this equilibrium.  When this condition is not met, a kinetic description is needed to capture the evolution of a distribution function in six-dimensional phase space (positions and velocities).

Weakly collisional plasmas require a kinetic description to capture a wealth of dynamical phenomena  that are wrapped up in the velocity distribution of particles.  From the perspective of a turbulence theorist, the following questions are particularly interesting: What are the dynamical invariants?  How do they travel through the phase space?  How do the velocity dimensions participate in the fully developed state of the turbulence?  Can one define an ``inertial range'' for the phase space in kinetic turbulence?  How can one, in general, adapt the phenomenology of fluid turbulence --- an energy cascade by local interactions, universal self-similar scaling, dissipation scale, and so forth?

In this paper we address these questions for a simple kinetic system: the two-dimensional electrostatic\footnote{The electrostatic limit in this context means that there are no magnetic fluctuations, although there is a strong mean magnetic field, constant in both space and time.} gyrokinetic system driven by a statistically homogeneous and isotropic source.  The gyrokinetic system of equations \cite[]{taylor-hastie, rutherford, catto1977, antonsen, frieman, brizard-hahm} is used to describe magnetised plasma dynamics on time scales much larger than the ion Larmor period.\footnote{Gyrokinetics has been developed by the magnetic fusion community to describe plasma turbulence that causes the transport of heat and particles in fusion devices.  A review of the theoretical framework of turbulence in magnetised plasmas is given by \cite{krommes}.  As demonstrated in the series of recent works by \cite{howes, howes3, schekochihin}, gyrokinetics is also appropriate for a wide range of astrophysical plasmas.}  In this limit, the presence of a strong, constant magnetic ``guide'' field causes the charged particles to rapidly trace nearly circular Larmor orbits so that the electrostatic fluctuations affect the motion adiabatically.  This causes slow drift motion perpendicular to the guide field.  Thus, what we are considering is essentially a kinetic theory of charged ``rings.''

We assume that the distribution function for the particles is a Maxwellian distribution plus a perturbation.  As a result, the intrinsic velocity scale in this problem is the thermal velocity, $\vth$.  Because of the magnetic guide field, this velocity scale gives rise to an associated intrinsic spatial scale which is $\rho$, the thermal Larmor radius $\rho = \vth/\Omega$, where $\Omega$ is the Larmor frequency.  Physically, the kinetic phenomenon on which we will be focusing arises as follows.  For electrostatic fluctuations on scales much larger than the Larmor radius, all particles move together with the same ${\bf E} \times {\bf B}$ drift and a fluid description is correct.  At the Larmor-radius scale and smaller, particles of distinct velocities have different effective ${\bf E} \times {\bf B}$ drifts (as they sample different regions of the electric field during their rapid Larmor motion) and a kinetic description is required.  At such scales, gyrokinetic turbulence exhibits a distinctive kinetic behaviour in phase space due to the presence of so-called ``nonlinear phase-mixing'' \cite[]{dorland-hammett-93, schek-ppcf, tatsuno-prl}.  Fluctuations in the distribution function nonlinearly cascade to create fine-scale structure in velocity space in addition to the two real-space dimensions --- thus, we may think of velocity space as an additional dimension to be treated on equal footing with position space.\footnote{Note that the nonlinear phase-mixing studied in this work depends critically on the assumption that the plasma is magnetised.  One should not necessarily expect the type of phase-space cascade considered here to appear in all types of kinetic turbulence.}

To address the problem of gyrokinetic turbulence, we borrow traditional methods from fluid turbulence.  This is possible because the gyrokinetic system bears some resemblance to equations of incompressible fluid turbulence.  The nonlinearity is convective, with a divergence-free velocity (the ${\bf E} \times {\bf B}$ drift velocity due to fluctuating electrostatic fields), while dissipation occurs via a collision operator that acts by second-order derivatives \cite[see ][]{catto1977, abel} in phase space and, therefore, acts on fine scales in analogy to viscosity in fluid turbulence.  In the long-wavelength cold-ion limit (see section \ref{chm-sec}), the electrostatic gyrokinetic system can be reduced to the Charney--Hasegawa--Mima (CHM) equation \cite[]{charney, hasegawa} or, with additional assumptions, to the inviscid vorticity equation that describes two-dimensional Euler turbulence \cite[]{kraichnan1967, batchelor}.\footnote{The relationship between plasma dynamics and two-dimensional fluid turbulence was demonstrated first by \cite{taylor}.}  Thus, from a fluid turbulence perspective, two-dimensional electrostatic gyrokinetics can be seen as a simple kinetic extension of extensively-studied fluid equations.

\subsection{Outline}

The progression of the paper is as follows.  We begin in section \ref{eqns-sec} by discussing the gyrokinetic system, its applicability and the specific assumptions and approximations we will be using.  In section \ref{invariants-sec}, we introduce and discuss the dynamical collisionless invariants of the gyrokinetic system.  It is shown that there are two such invariants.  One invariant is related to the perturbed free energy (or perturbed entropy) of the system.  We argue that this invariant must cascade forward to fine scales in phase space, where the collision operator ultimately dissipates it.  The other invariant is special to the two-dimensional electrostatic system and is not conserved in three-dimensional gyrokinetics.  We argue that this ``electrostatic'' invariant will cascade inversely to larger scales --- as does energy in two-dimensional fluid turbulence.

In section \ref{chm-sec}, we study the long wavelength regime ($k\rho \ll 1$) and explore the relationship between gyrokinetic and CHM/Euler turbulence.  We derive the inviscid CHM/Euler equations as asymptotic limits of the gyrokinetic system.  We also discuss the two invariants of the CHM/Euler equation and how they relate to the gyrokinetic collisionless invariants.  We find that our electrostatic invariant in this limit becomes the inversely cascading CHM/Euler invariant (`energy').  The forward cascading quantity from CHM/Euler turbulence (`enstrophy') is found to be only one part of the gyrokinetic quantity which cascades forward; the other essentially kinetic part corresponds to a passive scalar field.

In section \ref{phenom-sec}, we derive several scaling results by phenomenological arguments.  This section gives a simplified physical perspective, which may serve as a guide for the analysis in the remaining sections of the paper.  We begin with a standard treatment of the long-wavelength fluid regimes.  We then shift attention to the short-wavelength ($k\rho \gg 1$) nonlinear phase-mixing regime, where the turbulence has a strongly kinetic character --- this regime is the chief focus of the remainder of the paper.  We describe the physical mechanism of nonlinear phase-mixing and present a phenomenological analysis of the turbulence cascade based on the work by \cite{schek-ppcf, schekochihin}.  In section \ref{summary-cascades}, we tie together the results by presenting two scenarios -- injection at a long-wavelength scale and injection at a short-wavelength scale -- and describing the dual cascade through various scale ranges.

In section \ref{statistical-sec} we begin a more formal analysis of the gyrokinetic system.  We first introduce the statistical tools and notation that will be needed, followed by a discussion of symmetries.  Symmetries play an important role in turbulence theory, and in subsequent arguments we appeal to the principle of restored symmetry in the fully developed state.  In sections \ref{third-order-result-W-sec} and \ref{exact-inverse-sec}, we derive the third-order statistical laws following from the gyrokinetic equation, in the style of \cite{kolmogorov41c} and \cite{yaglom}.  These are exact results, valid uniformly for all wavenumbers, \ie both in the long- and short-wavelength limits.

In section \ref{spectral-theory-sec} we derive scaling laws for the phase-space spectra.  Scales in real space are treated in the conventional way by using a Fourier transform while for scales in velocity space, we use the zeroth-order Hankel transform, which is inspired by the mathematical structure of gyrokinetics.  Using this formalism, we derive approximate spectral scaling laws, both for the forward and inverse cascades, in the limit of spatial scales much smaller than the Larmor radius and velocity scales much smaller than the thermal velocity.

\subsection{Structure of paper}

\begin{figure}
\begin{center}
\includegraphics[width=\textwidth]{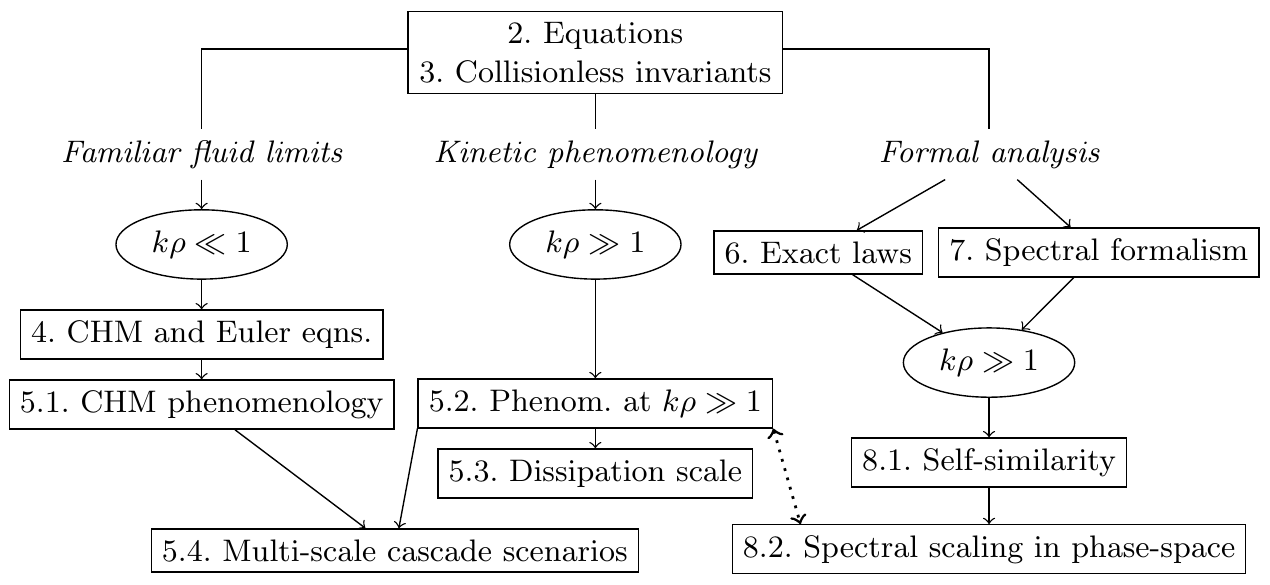}
\end{center}
\caption{Structure of the paper.  Unidirectional arrows indicate dependencies and flow of discussion.  Bidirectional arrows indicate cross-referencing.}
\label{paper-struc-fig}
\end{figure}

To help the reader navigate the paper, we have provided a diagram in figure \ref{paper-struc-fig} that describes its overall structure.  Although the paper is arranged to be read from start to finish, the reader may choose to narrow his/her attention by following only one of the three suggested paths in the diagram.  These paths are sufficiently independent to be read separately, but are strongly related.  The left path in the diagram focuses on the familiar CHM and Euler fluid limits in the context of gyrokinetic theory.  In the center path, many of the short-wavelength kinetic results are derived by simple phenomenological arguments.  The right path focuses on derivation of the results using exact methods and a novel spectral treatment of velocity space.  There is a significant amount of material in sections \ref{statistical-sec}, \ref{formalism-sec} and \ref{spectral-theory-sec} which is not covered in the phenomenological section \ref{phenom-sec}, but the derivations are more technical and the central path should provide a good introduction to the important ideas.

\section{Equations}\label{eqns-sec}

What follows is a brief discussion of the gyrokinetic equations, their normalisation and applicability.  A central point that emerges from this discussion is that the system studied in this paper will have wide-ranging applicability, subject to specific interpretation of the fluctuating fields and normalisation.  Readers wishing to skip these details are advised to take equations \ref{gyro-g} and \ref{qn-g} with definitions \ref{gyro-avg-def}, \ref{vE-def-eqn} and \ref{angle-avg-def} as the equations we will study in this paper.  

In kinetic plasma theory, each species $s$ (electrons or ions) of a plasma is described by a distribution function $f$, which is a function of space, time and velocity.  As is customary with so-called $\delta f$ gyrokinetics, the full distribution function is split into an equilibrium part and fluctuating parts:

\begin{equation}
f = F_{0} + \epsilon\left[- \phi F_{0} + h(t, {\bf R}, v_{\perp}, v_{\parallel})\right],\label{f-pert-expan}
\end{equation}

\noindent where $\phi$ is the electrostatic potential and $\epsilon = \rho/L$ is the fundamental gyrokinetic expansion parameter, the ratio of the Larmor radius to the system scale.  The equilibrium distribution $F_0 = (2\pi)^{-3/2}\e^{-v^2/2}$ is a Maxwellian distribution in velocity space.  The second term in equation \ref{f-pert-expan} is the so-called Boltzmann part of the distribution.  The final term is the gyro-centre distribution function, $h$, which depends on coordinates that are defined relative to the mean magnetic field, ${\bf B}_0 = {\bf \hat{z}} B_0$: the perpendicular velocity $v_{\perp}$ (magnitude of the velocity perpendicular to ${\bf B}_0$), the parallel velocity $v_{\parallel}$ (velocity parallel to ${\bf B}_0$) and the gyro-centre position ${\bf R}$.  The particle position and gyro-centre position are related by the transformation ${\bf r} = {\bf R} + \rhovec$ where the Larmor radius vector is $\rhovec(\vartheta) = {\bf {\hat z}}\times{\bf v} = v_{\perp}({\bf {\hat y}}\cos{\vartheta} - {\bf {\hat x}}\sin{\vartheta})$ and $\vartheta$ is the gyro-angle.  Velocities are normalised to the thermal speed $\vth$, while spatial scales are normalised to the Larmor radius $\rho = \vth/\Omega$.  The normalisations are given in full in the appendix \ref{norm-sec}.

It can be shown by rigourous asymptotic expansion \cite[\eg, see][]{howes} that the gyro-centre distribution function $h$ satisfies the nonlinear gyrokinetic equation.  The electrostatic gyrokinetic equation for a homogeneous equilibrium (\ie, assuming no variation in the background magnetic field, density or temperature) is

\begin{equation}
\frac{\partial h}{\partial t} + v_{\parallel} \frac{\partial h}{\partial z} + {\bf v}_E\cdot\bnabla h - \CollisionOp{h} = \frac{\partial\gyroavg{\phi}}{\partial t} F_0 + \Source,%-  {\bf v}_E \cdot {\bf {\hat x}} F_0
\label{gyro_full}
\end{equation}

\noindent where $\bnabla = \partial/\partial{\bf R}$.  The term $\CollisionOp{h}$ represents the gyrokinetic collision operator, and $\Source$ is a general kinetic source term (\ie external drive), which will be left unspecified at this stage.  The gyro-average $\gyroavg{.}$ is an average over the gyro-angle, with gyro-centre position ${\bf R}$ held fixed: for an arbitrary function of position $A({\bf r})$, 

\begin{equation}
\gyroavg{A({\bf r})} = \frac{1}{2\pi}\int_0^{2\pi} d\vartheta A({\bf R} + \rhovec(\vartheta)).
\label{gyro-avg-def}
\end{equation}

\noindent Note that, even if $A({\bf r})$ has no velocity dependence, its gyro-average $\gyroavg{A({\bf r})}$ does.  The gyro-averaged ${\bf E} \times {\bf B}$ velocity, a function of gyro-centre position ${\bf R}$ and velocity $v_{\perp}$, is defined

\begin{equation}
{\bf v}_E({\bf R}, v_{\perp}) = {\bf {\hat z}}\times\bnabla\gyroavg{\phi}.\label{vE-def-eqn}
\end{equation}

In our two-dimensional theory, we neglect the parallel ion inertia term $v_{\parallel} \partial h/\partial z$ in equation \ref{gyro_full} (see section \ref{future-work-sec} for a discussion of this).  The $v_{\parallel}$ dependence of $h$ is therefore no longer important (except for its appearance in the collisional operator\footnote{When written explicitly, the full collisional operator is the gyro-averaged linearised Landau operator \cite[see the appendix of][]{schekochihin}, an integro-differential operator with non-trivial dependence on $v_{\parallel}$ space.  However, a simplified operator, such as the one given by \cite{abel}, is a good example to keep in mind for the discussions of this paper.}).  Thus, it will be concealed when possible in the remainder of this paper.  It should be assumed that wherever velocity integration is present, integration over $v_{\parallel}$ is implied.  Henceforth, we use $v_{}$ as shorthand for the perpendicular velocity $v_{\perp}$.  

We can rewrite the gyrokinetic equation \ref{gyro_full} compactly in terms of a new distribution function $g = h - \gyroavg{\phi}F_0$.  This is two-dimensional gyrokinetic equation that we will use in the rest of this paper:

\begin{equation}
\frac{\partial g}{\partial t} +  {\bf v}_E \cdot\bnabla g = \CollisionOp{h} + \Source.
\label{gyro-g}
\end{equation}

The full plasma dynamics are described by the gyrokinetic equations for each species, with the additional constraint that charge neutrality must be maintained (``quasi-neutrality'').  We make a common simplification in solving the gyrokinetic equation for only one of the species.  This is justified by the disparity between the ion and electron Larmor radii, which leads to the Boltzmann (or `adiabatic') response for the other species (see appendix \ref{boltz-sec} for a discussion).  Under these assumptions, quasi-neutrality takes the form

\begin{equation}
2\pi\int_0^{\infty} v dv \angleavg{g} = (1 + \tau)\phi - \Gamma_0\phi.
\label{qn-g}
\end{equation}

\noindent The constant $\tau$ depends on the choice of kinetic species (or scale of interest) and response model for the non-kinetic species (see appendix \ref{boltz-sec}).  Note that the velocity integration at constant real-space coordinate ${\bf r}$ requires a change of variables (from gyro-center coordinate {\bf R}).   This gives rise to the angle average $\angleavg{.}$, which is an average over gyro-angle $\vartheta$ with real-space coordinate ${\bf r}$ held fixed:

\begin{equation}
\angleavg{A({\bf R})} = \frac{1}{2\pi}\int_0^{2\pi} d\vartheta A({\bf r} - \rhovec(\vartheta)).
\label{angle-avg-def}
\end{equation}

\noindent The operator $\Gamma_0$ is defined 

\begin{equation}
\Gamma_0\phi = \int_0^{\infty} v dv \;\e^{-v^2/2}\angleavg{\gyroavg{\phi}}.\label{gamma-0-def}
\end{equation}

\noindent This operator is multiplicative in Fourier space --- it is simply 

\begin{equation}
\hat{\Gamma}_0(k) = I_0(k^2)e^{-k^2},\label{gamma-hat-def}
\end{equation}

\noindent where $I_0$ is the zeroth-order modified Bessel function.  Note that the gyro-average in Fourier space reduces to multiplication by the Bessel function $J_0(k v_{})$.  So, equation \ref{vE-def-eqn} is

\begin{equation}
\hat{\bf v}_E = \i {\bf {\hat z}}\times{\bf k}J_0(k v_{})\hat{\phi}({\bf k}),\label{vE-def-eqn-fourier}
\end{equation}

\noindent where $\hat{\phi}$ denotes the Fourier transform of $\phi$.  Quasi-neutrality, equation \ref{qn-g}, in Fourier space becomes

\begin{equation}
\hat{\phi} ({\bf k}) = \beta(k)\int v_{}dv_{} J_0(k v_{}) \hat{g}({\bf k}, v_{})\label{qn-g-k-1},
\end{equation}

\noindent where

\begin{equation}
\beta(k) = \frac{2\pi}{(1 + \tau) - \hat{\Gamma}_0(k)}. \label{beta-def-eqn}
\end{equation}

Equations \ref{gyro-g} and \ref{qn-g} are the starting point for all further investigations of this paper.  This two-dimensional gyrokinetic system is the simplest paradigm for kinetic turbulence of magnetised plasmas.  

\section{Collisionless invariants}\label{invariants-sec}

We focus on two quadratic invariants of the two-dimensional gyrokinetic system.  To derive the first, let us start by noting that the average of $g^2$ is collisionlessly conserved for each individual value of $v_{}$.  That is, the volume-averaged $g^2$, as a function of velocity, must remain unchanged in the absence of collisions or forcing.  Dividing $g^2$ by an arbitrary function of velocity, $\kappa(v)$, and integrating over $v$, a class of invariants is formed, which we will refer to as the ``generalized free energy'' (after this section, we use the abbreviation ``free energy''):

\begin{equation}
W_g[\kappa(v)] = 2\pi\int v dv\int \frac{d^2{\bf R}}{V} \frac{g^2}{2\kappa(v)} = 2\pi\int v dv \int \frac{d^2{\bf r}}{V}\frac{\angleavg{g^2}}{2\kappa(v)},\label{gen-free-energy-0}
\end{equation}

\noindent where $V$ is the system volume.  A scaling theory for two-dimensional gyrokinetic turbulence may be developed, while allowing for arbitrary $\kappa(v)$ --- under the constraint that $\kappa(v)$ varies only on the thermal velocity scale $\vth = 1$ and so does not introduce additional velocity-space scales into the problem.  Two functions, $\kappa = 1$ and $\kappa = F_0$, are particularly useful, so we introduce the following notation:

\begin{subequations}
\label{free-energy}
\begin{align}
W_{g1} &= W_g[1], \label{free-energy-a}\\
W_{g0} &= W_g[F_0].\label{free-energy-b}
\end{align}
\end{subequations}
 
The second invariant of the gyrokinetic system, equations \ref{gyro-g} and \ref{qn-g}, is particular to the electrostatic two-dimensional case:

\begin{equation}
E_{} = \frac{1}{2}\int \frac{d^2{\bf r}}{V}\left[(1 + \tau)\phi^2 - \phi\Gamma_0\phi\right].\label{E-def}
\end{equation}

\noindent We refer to this quantity as the ``electrostatic invariant.''  Note that the operator $\Gamma_0$ renders the second term negligible at large $k$ (because in Fourier space $\hat{\Gamma}_0 \sim k^{-1}$; see equation \ref{gamma-hat-def}).  For this and other reasons, it is convenient to define the quantity $E_{\phi}$ (which is not in general an invariant):

\begin{equation}
E_{\phi} = \frac{1}{2}\int \frac{d^2{\bf r}}{V}\; \phi^2 \label{E-phi-def}.
\end{equation}

\noindent From equation \ref{E-def}, $E_{}(k) \approx (1+ \tau)E_{\phi}(k)$ for $k \gg 1$.  

We now demonstrate the conservation of $W_g$ and $E_{}$ by deriving the ``global budget'' equations from the gyrokinetic system.  Multiplying the gyrokinetic equation \ref{gyro-g} by $g_{}/\kappa$, averaging over gyro-centre position {\bf R} and integrating over velocity, we get

\begin{equation}
\frac{d W_g}{d t} = \int \frac{d^2{\bf r}}{V} 2\pi\int vdv \angleavg{\frac{g_{}\CollisionOp{h}}{\kappa(v)}} + \varepsilonW,\label{global-budget-free-energy-0}
\end{equation}

\noindent where the injection rate of free energy is

\begin{equation}
\varepsilonW = \int \frac{d^2{\bf r}}{V} 2\pi\int vdv \frac{\angleavg{g \Source}}{\kappa}.
\end{equation}

Multiplying the gyrokinetic equation \ref{gyro-g} by $\gyroavg{\phi}$ and integrating over real-space coordinate {\bf r} and velocity $v$, we obtain the global budget equation for the electrostatic invariant:

\begin{equation}
\frac{d E}{dt} = \frac{1}{2}\int \frac{d^2{\bf r}}{V}\phi\; 2\pi\int vdv \angleavg{\CollisionOp{h}} + \varepsilonE,
\end{equation}

\noindent where the injection rate of the electrostatic invariant is

\begin{equation}
\varepsilonE = \frac{1}{2}\int \frac{d^2{\bf r}}{V} \phi \; 2\pi \int v dv \angleavg{\Source}.
\end{equation}

It is useful to compare these results with what is known for three-dimensional gyrokinetics.  Note that the electrostatic invariant is not conserved by the three-dimensional gyrokinetic equation \ref{gyro_full}, because of the particle streaming term $v_{\parallel}\partial h/\partial z$ that must be retained (just as the three-dimensional Navier--Stokes equation does not conserve enstrophy, while in two dimensions it does).  Three-dimensional gyrokinetics does have an invariant that plays the role of the Navier--Stokes energy.  The invariant is the perturbed free energy, $W_{}$ \cite[see][]{krommes-hu, sugama96, howes, schek-ppcf, schekochihin}.  Its conservation is demonstrated by multiplying the gyrokinetic equation \ref{gyro_full} by $h/F_0$ and integrating over velocity and position ${\bf r}$.  In our notation, $W$ is

\begin{equation}
W_{} = W_{g0} + E_{}.
\end{equation}

\noindent For scales much smaller than the Larmor radius, we will see that the electrostatic invariant contributes negligibly to $W_{}$ and it is therefore justified to refer to either $W_{}$ or $W_{g0}$ as the free energy.

\section[CHM and Euler eqns.]{Charney--Hasegawa--Mima and Euler equations}\label{chm-sec}

The CHM and Euler equations are popular paradigms for two-dimensional fluid turbulence.  They can be derived as rigourous (and physical) asymptotic limits of the gyrokinetic equation, as we now demonstrate.

We begin with the gyrokinetic system \ref{gyro-g} and \ref{qn-g}, and take the long-wavelength limit, $k \ll 1$, and the small temperature ratio limit $\tau \ll 1$.\footnote{The limit of small temperature ratio is somewhat restrictive but the CHM equation follows from it in a simple and transparent manner.  We note that a more detailed system of fluid equations can be worked out for $\tau$ of order unity but, for our purposes, it is not substantially different from the familiar CHM equation.}  Formally, we assume $k^2 \sim \tau$ and $\partial_t \sim \CollisionOpZ{.} \sim k^2 \phi$.  The gyro-average in equation \ref{gyro-g} acts as unity, while we must keep $\Gamma_0$ to order $k^2$ in equation \ref{qn-g} (the zeroth-order part cancels), so $\Gamma_0 \approx 1 + \nabla^2$.  The gyrokinetic system in this limit is

\begin{subequations}\label{gyro-sys-chm}
\begin{gather}
\frac{\partial g}{\partial t} +   {\bf v}_{E0}\cdot\bnabla g = \CollisionOpZ{g} + \Source,\label{gyro-chm} \\ 
2\pi\int v_{} dv_{} g =  \tau\phi - \nabla^2\phi \label{qn-chm},
\end{gather}
\end{subequations}

\noindent where ${\bf v}_{E0} = {\bf {\hat z}}\times\bnabla\phi$ is the ${\bf E} \times {\bf B}$ velocity (without the gyro-average).  Now, by integrating equation \ref{gyro-chm} over velocity space (noting that the integral of the collision operator is zero by particle conservation) and using quasi-neutrality, equation \ref{qn-chm}, we obtain a single equation for the electrostatic potential:

\begin{equation}
\partial_t(\tau - \nabla^2)\phi + {\bf v}_{E0}\cdot\bnabla (-\nabla^2\phi) = \SourceCHM, 
\label{chm-eqn}
\end{equation}

\noindent where $\SourceCHM = 2\pi\int vdv \Source$.  This is the inviscid CHM equation.  The scale given by $\tau^{-1/2}$ corresponds to the Rossby deformation radius in quasi-geostrophic turbulence or the ion sound radius, $\rho_s$, in a plasma.  The two-dimensional Euler equation is obtained in the case $\tau = 0$ (the no-response model; see appendix \ref{boltz-sec}, equation \ref{tau-def}).

There are two invariants of the CHM/Euler equation that are relevant to our discussion.  These are referred to as energy and enstrophy, although their physical interpretation depends on the specific scale of interest.  They are found by multiplying equation \ref{chm-eqn} by $\phi$ and $\nabla^2\phi$, respectively, and integrating over the system volume:

\begin{subequations}
\label{chm-invar-defs}
\begin{align}
& \ECHM = \frac{1}{2}\int \frac{d^2{\bf r}}{V}(\tau\phi^2 + |\bnabla\phi|^2), \\
& \ZCHM = \frac{1}{2}\int \frac{d^2{\bf r}}{V}(\tau|\bnabla\phi|^2 + (\nabla^2\phi)^2).
\end{align}
\end{subequations}

\noindent We now show that there is an additional invariant associated with the gyrokinetic description (given by equation \ref{gyro-sys-chm}).  Without loss of generality, we introduce the following {\em ansatz} for $g$:

\begin{equation}
g = F_0 (\tau - \nabla^2)\phi + \tilde{g}, \label{g-ansatz}
\end{equation}

\noindent Quasi-neutrality (equation \ref{qn-chm}) implies that $\tilde{g}$ does not contribute to particle density:

\begin{equation}
\int v dv\; \tilde{g} = 0.
\end{equation}

\noindent Now, multiplying the CHM equation \ref{chm-eqn} by $F_0$ and subtracting it from equation \ref{gyro-chm}, we obtain a separate equation for $\tilde{g}$:

\begin{equation}
\frac{\partial \tilde{g}}{\partial t} + {\bf v}_{E0} \cdot\bnabla \tilde{g} = \CollisionOp{\tilde{g}} + (\Source - F_0\SourceCHM). \label{gyro-twiddle}
\end{equation}

\noindent Thus, $\tilde{g}$ behaves as a passive scalar in the CHM/Euler limit, advected by the ${\bf E} \times {\bf B}$ flow determined by equation \ref{chm-eqn}.  From equation \ref{gyro-twiddle}, it follows that there is another collisionless invariant:

\begin{equation}
W_{\tilde{g}}[\kappa(v)] = \frac{1}{2}\int\frac{d^2{\bf r}}{V}\int v dv \frac{\tilde{g}^2}{\kappa(v)},\label{w-g-twiddle-def}
\end{equation}

\noindent where $\kappa(v)$ is an arbitrary function of velocity.  The free energy invariant $W_{g0}$ (equation \ref{free-energy-b}) can now be written in terms of the three gyrokinetic--CHM invariants:

\begin{equation}
W_{g0} = W_{\tilde{g}0} + \tau \ECHM + \ZCHM, \label{invar-reln}
\end{equation}

\noindent where we define $W_{\tilde{g}0} = W_{\tilde{g}}[F_0]$.  As we have shown, each term here is conserved individually in the CHM/Euler limit.

To complete our account of how the gyrokinetic invariants reduce to the CHM/Euler invariants in the long-wavelength limit, we note that the (gyrokinetic) electrostatic invariant $E_{}$ reduces to the energy invariant $\ECHM$.  From equation \ref{E-def}, taking $\Gamma_0 \approx 1 + \nabla^2$, we have

\begin{equation}
E = \ECHM.\label{E-ECHM-reln}
\end{equation}

As a final comment in this section, note that viscosity did not appear in the fluid limit derived above.  To recover the viscous CHM and the two-dimensional Navier--Stokes equation, one must raise the ordering of the collisional operator.  Specifically, the viscous term can be obtained by ordering $\CollisionOpZ{.} \sim k^{-2}\partial_t$.

\section{Phenomenology}
\label{phenom-sec}

In sections \ref{statistical-sec}, \ref{formalism-sec} and \ref{spectral-theory-sec}, we will formally derive properties of fully developed gyrokinetic turbulence.  In this section, we give a simpler derivation of some of the scaling results via a set of phenomenological arguments in the style of \cite{obukhov41a, obukhov41b}.  We first sketch the phenomenology for stationary, driven CHM turbulence, and then outline how these arguments can be adapted for the essentially kinetic short-wavelength regime.

\subsection[CHM phenomenology]{CHM range, $k \ll 1$}\label{chm-phenom-sec}

Inspection of equation \ref{chm-invar-defs} reveals that the two invariants of the CHM and Euler equations have spectra that are constrained to satisfy $\ZCHM(k)/\ECHM(k) = k^2$.  By the argument due to \cite{fjortoft}, this implies that a dual cascade will occur.  That is, injection at a scale $k_i$ will give rise to two cascades: an inverse cascade of energy to $k < k_i$ and a forward cascade of enstrophy to $k > k_i$.

Using the assumptions of isotropy and local (in scale) interactions, the CHM equation \ref{chm-eqn} is translated into phenomenological language as follows:

\begin{equation}
\omegaNL (\tau + \ell^{-2})\varphi_{\ell} \sim \varphi_{\ell}^2\ell^{-4},\label{chm-phenom-eqn}
\end{equation}

\noindent where $\omegaNL$ is the nonlinear decorrelation rate and $\varphi_{\ell}$ is the characteristic amplitude of $\phi$ at scale $\ell \sim k^{-1}$.  The essential assumption that leads to a scaling theory is that the flux of enstrophy is constant for $k > k_i$, while the flux of energy is constant for $k < k_i$.  Multiplying the right hand side of equation \ref{chm-phenom-eqn} by $\ell^{-2}\phi_{\ell}$ gives the nonlinear flux of enstrophy, which must be equal to the total rate of enstrophy injection.  This flux must be independent of scale in the stationary state: $\ell^{-6}\phi_{\ell}^3 \sim \varepsilon_Z = \mbox{const.}$  Therefore,

\begin{equation}
\varphi_{\ell} \sim \varepsilon_Z^{1/3}\ell^2\label{phi-forward-scaling-chm}
\end{equation}

\noindent and the spectrum of potential fluctuations\footnote{The spectra $\ECHM(k)$ and $\ZCHM(k)$ have different scaling laws depending on the relative sizes of $k^2$ and $\tau$.  Thus, it is more convenient to speak about the spectrum $E_{\phi}(k)$ (defined by equation \ref{E-phi-def}), which has a single scaling law.} is $E_{\varphi}(k) \sim k^{-5}$.  

To find the scaling law for the inverse-cascade range, $k < k_i$, we multiply the right-hand side of equation \ref{chm-phenom-eqn} by $\varphi_{\ell}^2$ to get the energy flux, and set this equal to the energy injection: $\ell^{-4}\phi_{\ell}^3 \sim \varepsilonE = \mbox{const}$.  This implies 

\begin{equation}
\varphi_{\ell} \sim \varepsilonE^{1/3}\ell^{4/3}\label{chm-inverse-phi-scaling}
\end{equation}

\noindent and, therefore, $E_{\phi}(k) \sim k^{-11/3}$.
  
Finally, the cascade of $W_{\tilde{g}}$ (equation \ref{w-g-twiddle-def}) is passive and directed forward.  Constancy of nonlinear flux of this invariant implies $\ell^{-2}\tilde{g}^2_{\ell}\phi_{\ell} \sim \tilde{\varepsilon} = \mbox{const.}$\footnote{Note that the passive cascade at $k > k_i$ is marginally non-local because the decorrelation rate scales as $\omegaNL \sim \varphi_{\ell}\ell^{-2}$, which, by equation \ref{phi-forward-scaling-chm}, implies that all scales in the forward-cascade range contribute similarly to the time evolution of $\tilde{g}$.}  Using equation \ref{phi-forward-scaling-chm}, we have for $k > k_i$

\begin{equation}
\tilde{g}_{\ell}  \sim \tilde{\varepsilon}^{1/2}\varepsilon_Z^{-1/6} = \mbox{const.}.
\end{equation}

\noindent Therefore, the spectrum of fluctuations of $\tilde{g}$ scales as $W_{\tilde{g}}(k) \sim k^{-1}$, the usual \cite{batchelor-passive} scaling for a passive scalar \cite[also, see][for the two-dimensional case]{lesieur}.

\subsection[Phenom. at $k\rho \gg 1$]{Nonlinear phase-mixing range, $k \gg 1$}\label{nl-mix-range-phenom-sec}

\cite{schek-ppcf, schekochihin} proposed a phenomenological description of the large-$k$ forward cascade in gyrokinetic turbulence.  We reproduce their arguments here, in a slightly different form, and also propose a phenomenology to describe the inverse cascade.  This will motivate the formal derivations of sections \ref{statistical-sec} and \ref{spectral-theory-sec} and highlight the salient physical processes.

\begin{figure}
\begin{center}
\includegraphics[width=.7 \columnwidth]{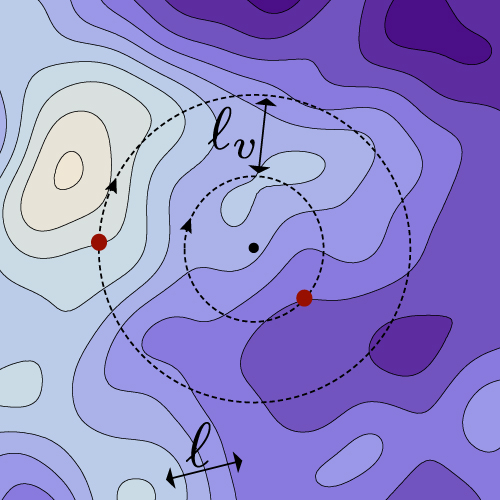}
\end{center}
\caption{Correlation in velocity space.  Contours of constant potential are plotted in a blue color scheme.  Overlaid are the orbits of two particles that have the same gyro-centre.  During a nonlinear turnover time, the phase-mixing process will de-correlate the motion of these particles because their Larmor radii differ by the correlation length $\ell$ of the potential.  Thus, $\ell_v \sim \ell$.  (Colour online.)}
\label{v-correl-fig}
\end{figure}

In gyrokinetics, the drift motion of particles is determined by fluctuating fields averaged along the particle Larmor orbits.  Two particles sharing the same gyro-centre but having different velocities perpendicular to the equilibrium magnetic field are subject to a different averaged electrostatic potential and thus have distinct ${\bf E} \times {\bf B}$ drifts.  This is illustrated in figure \ref{v-correl-fig}.  One can see from this cartoon that if the ${\bf E} \times {\bf B}$ flow has a correlation length $\ell$, then the particle motion must be decorrelated over the velocity-space scale $\ell_v$ such that $\ell_v \sim \ell$.

Now let $g_{\ell}(v)$ be the characteristic amplitude of the distribution function $g$ at scale $\ell$.  In view of the above argument, we assume this quantity to have random structure in velocity space on the scale $\ell_v \sim \ell$.  From quasi-neutrality (equation \ref{qn-g}) we can estimate the scaling of $\phi_{\ell}$ in terms of $g_{\ell}$.  We make the substitution $k^{-1} \rightarrow \ell$ in the Fourier-space expression \ref{qn-g-k-1} and take $k \gg 1$ (so that $\beta(k)$ is a constant).  The relationship between $\phi_{\ell}$ and $g_{\ell}$ is then derived as follows:

\begin{equation}
\phi_{\ell} \sim \int v dv J_0(v/\ell)g_{\ell}(v) \sim \ell^{1/2}\int v^{1/2} dv \cos(v/\ell - \pi/4)g_{\ell}(v) \sim \ell^{1/2}\ell_v^{1/2} g_{\ell},\label{phi-g-ell-scaling-reln}
\end{equation}

\noindent where we have used the large-argument approximation of the Bessel function $J_0$.  Because of the random velocity dependence of $g_{\ell}(v)$, we argue that the velocity integral accumulates like a random walk --- the number of steps scales as $\ell_v^{-1}$ and the step-size scales as $\ell_v$, so the typical ``displacement'' scales as $\ell_v^{1/2}$.  Note that in the final expression we have introduced the amplitude $g_{\ell}$ (distinct from $g_{\ell}(v)$), which can roughly be interpreted as the root-mean-square value of the fluctuations in $g_{\ell}(v)$ over velocity space: $g_{\ell} \sim \sqrt{\int v dv g_{\ell}^2(v)}$.  Taking $\ell_v \sim \ell$ (as motivated by the arguments illustrated in figure \ref{v-correl-fig}) in equation \ref{phi-g-ell-scaling-reln}, we find

\begin{equation}
\phi_{\ell} \sim \ell g_{\ell}.
\label{phi-accum}
\end{equation}

This immediately implies that there is a possibility of a dual cascade for the two invariants $W_g$ and $E_{}$.  Indeed, if \ref{phi-accum} holds, the spectra $W_g(k)$ and $E(k)$ must satisfy $W_g(k) \sim k^2 E(k)$.  By the argument of \cite{fjortoft}, this constraint leads to a dual cascade where $E_{}$ is transferred to larger scales and $W_g$ is transferred to smaller scales.  Thus, if the injection scale is $k_i \gg 1$, we predict the existence of two cascades: a forward cascade of $W_g$ to $k > k_i$ and an inverse cascade of $E_{}$ to $k < k_i$.

From the gyrokinetic equation \ref{gyro-g}, we estimate the nonlinear decorrelation rate from the nonlinear term by substituting ${\bf v}_E\cdot\bnabla \rightarrow J_0(v/\ell)\phi_{\ell}\ell^{-2}$, where the Bessel function comes from the gyro-average in equation \ref{vE-def-eqn-fourier}.  Then the nonlinear decorrelation rate is

\begin{equation}
\label{flux-phenom}
\omegaNL \sim \ell^{-3/2}\phi_{\ell},
\end{equation}

\noindent where we have used the large argument approximation for $J_0(v/\ell)$ to estimate that the gyro-averaging introduces a factor of $\ell^{1/2}$.  To determine scalings in the forward-cascade range, we assume locality of interactions and constancy of the flux of free energy (see section \ref{statistical-sec}, equation \ref{four-fifths-iso} for the exact statement of constancy of flux).  This gives

\begin{equation}
\omegaNL g_{\ell}^2 \sim \ell^{-3/2}\phi_{\ell}g_{\ell}^2 \sim \varepsilonW = \mbox{const.}
\end{equation}
 
\noindent Using equation \ref{phi-accum}, we have $g_{\ell}^3\ell^{-1/2} \sim \varepsilonW$ and, therefore, 

\begin{align}
&g_{\ell} \sim \varepsilonW^{1/3} \ell^{1/6},\label{g-ell-scaling-phenom}\\
&\phi_{\ell} \sim \varepsilonW^{1/3} \ell^{7/6}.\label{phi-scaling-phenom}
\end{align}
 
\noindent The resulting spectra are $W_g(k) \propto k^{-4/3}$ and $E(k) \propto k^{-10/3}$ (in section \ref{spectral-scaling-sec} these are derived more formally).  These spectra have been confirmed numerically by \cite{tatsuno-prl, tatsuno-pop}.

The analysis of the inverse-cascade range, $k < k_i$, follows as in section \ref{chm-phenom-sec}.  Assuming constancy of the flux of the electrostatic invariant $E_{}$ in the inverse-cascade range (see section \ref{exact-inverse-sec}, equation \ref{four-fifths-E-iso}), we have, using equation \ref{flux-phenom},

\begin{equation}
\omegaNL \varphi_{\ell}^2 \sim \ell^{-3/2} \phi_{\ell}^3 \sim \varepsilon_E.
\end{equation}

\noindent Hence

\begin{equation}
\phi_{\ell} \sim \varepsilonE^{1/3}\ell^{1/2}.\label{phi-scaling-phenom-inverse}
\end{equation}

\noindent The corresponding spectrum is $E(k) \sim E_{\phi}(k) \sim k^{-2}$.  If we assume that the relationship \ref{phi-accum} also holds in the inverse-cascade range, equation \ref{phi-scaling-phenom-inverse} implies

\begin{equation}
g_{\ell} \sim \varepsilonE^{1/3}\ell^{-1/2}\label{g-ell-scaling-phenom-inverse}
\end{equation}

\noindent and so the free-energy spectrum in this range is $W_g(k) \sim k^0$.  See section \ref{inverse-cascade-sec} for a more formal derivation of these scalings.

We note that the numerical experiments of \cite{tatsuno-jpf} have confirmed that the locality of the cascade is a good assumption and that the relationship $\ell_v \sim \ell$ is well satisfied for both the forward and inverse cascades.

\subsection[Dissipation scale]{Dissipation scale}\label{dissip-scale-sec}

Having derived the inertial-range scalings for the forward cascade, it is easy to estimate the collisional cutoff scale that marks the end of the inertial range.  For this purpose, we notice that the collision operator acts by second derivatives in phase space, and so assuming $\ell_v \sim \ell$, we may take the collisional damping rate at scale $\ell$ to be $\omega_{\mbox{\scriptsize{coll}}} \sim \nu \ell^{-2}$, where $\nu$ is the collision frequency.  At the collisional cutoff scale $\ell_c$, the nonlinear and dissipation rates must balance: $\omegaNL(\ell_c) \sim \omega_{\mbox{\scriptsize{coll}}}(\ell_c)$.  Using equations \ref{flux-phenom} and \ref{phi-scaling-phenom}, we obtain

\begin{equation}
\ell_c \sim k_c^{-1} \sim \left(\frac{\nu^3}{\varepsilonW}\right)^{1/5},\label{dissip-scale-eqn}
\end{equation}

\noindent which is the gyrokinetic analog of the Kolmogorov scale.  Note that the cutoff scale may be written in terms of ${\rm Do}$, the Dorland number \cite[][]{tatsuno-prl}.  Setting $\ell_{} = 1$ in equation \ref{phi-scaling-phenom}, one finds $\varepsilonW \sim \phi_1^3$ and so $k_c \sim {\rm Do}^{3/5}$, where ${\rm Do} = \varphi_1/\nu$.  ${\rm Do}$ plays the role of the Reynolds number for the nonlinear phase-space cascade in gyrokinetic turbulence, where the outer scale of the turbulence is taken to be the Larmor scale because it marks the start of the nonlinear phase-mixing range.\footnote{In dimensional units, ${\rm Do} = \omegaNL(\rho)/\nu =  \varphi_{\rho}/B_0 \rho^2\nu$, where $\omegaNL(\rho)$ is the nonlinear decorrelation rate and $\varphi_{\rho}$ is the fluctuation amplitude, both evaluated at the Larmor scale.}

\subsection[Multi-scale cascade scenarios]{Gyrokinetic cascade through multiple scales}\label{summary-cascades}

Using the phenomenological scalings derived above, we now give a sketch of the fully-developed cascade from the low-$k$ CHM/Euler limit to the high-$k$ nonlinear phase-mixing limit.  Note that the kinetic invariants $W_g$ and $E_{}$ are exact invariants for both the high-$k$ and low-$k$ regimes, so they cascade through this entire range, including $k \sim 1$ (there is no dissipation at the Larmor scale).  We may use what we have learned about the relationship between the CHM/Euler equations and our two-dimensional gyrokinetic system to describe this cascade in detail.  In particular, equation \ref{invar-reln} states that the gyrokinetic invariant $W_g$ is composed of three separately conserved quantities in the CHM limit.  In spectral form, the equation is

\begin{equation}
 W_{g}(k) = W_{\tilde{g}}(k) + \tau \ECHM(k) + \ZCHM(k) = W_{\tilde{g}}(k) + \left[\tau^2 + 2\tau k^2 + k^4\right]E_{\varphi}(k).\label{Wg-CHM-spec-reln}
\end{equation}

\noindent For the electrostatic invariant, equation \ref{E-ECHM-reln} gives

\begin{equation}
E(k) \approx \ECHM(k) = (\tau + k^2)E_{\varphi}(k).\label{E-ECHM-spec-reln}
\end{equation}

The cascades traverse two special physical scales, the ion Larmor radius ($k = 1$) and the ion sound radius ($k = \tau^{1/2}$).  While nothing is dissipated at these scales, the physics of the turbulent fluctuations changes, giving rise to different scaling regimes.  

The range $k \ll \tau^{1/2}$ is referred to as the `potential limit' of the CHM equation, because the spectrum of $\ECHM$ becomes proportional to the potential fluctuation spectrum, $E_{\varphi}(k)$.  The range $\tau^{1/2} \ll k \ll 1$ is called the `kinetic limit' of the CHM equation because $\ECHM$ may be interpreted as the kinetic energy density of the flow.  The range $1 \ll k \ll k_c$ (where $k_c$ is the dissipation scale; see section \ref{dissip-scale-sec}) defines what we have been referring to as the nonlinear phase-mixing regime.

We have assumed that there is localised injection of free energy at a single wavenumber  $k_i$.  We now examine two possible scenarios: $k_i \ll \tau^{1/2}$ (injection at large scales) and $k_i \gg 1$ (injection at small scales), depicted in figures \ref{large-inject-fig} and \ref{small-inject-fig}, respectively.

\subsubsection{Injection at large scales}
  
\begin{figure}
\begin{center}
\includegraphics[width=\columnwidth]{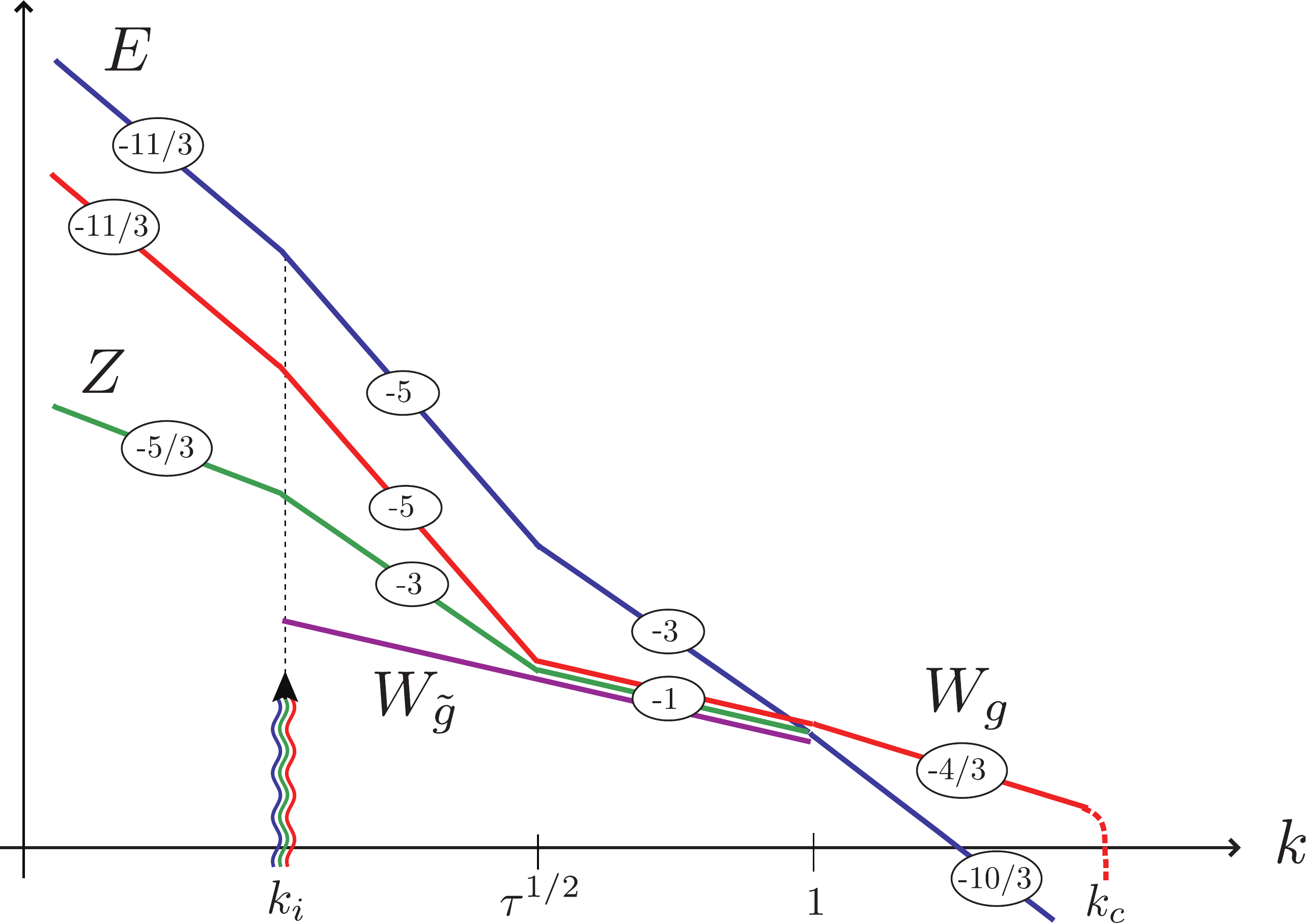}
\end{center}
\caption{Scenario 1: Injection at large scales.  The scaling indices for each sub-range are printed in bubbles over the plots of the invariants.  (Colour online.)}
\label{large-inject-fig}
\end{figure}

In figure \ref{large-inject-fig}, the injection is at large scales, in the potential limit of the CHM equation, where $k_i \ll \tau^{1/2}$.  In this limit, the electrostatic invariant $E_{}$ reduces to $E_{} \approx \tau E_{\phi}$.  An inverse cascade of the electrostatic invariant occurs at $k \ll k_i$, with the scaling derived in the section \ref{chm-phenom-sec} (equation \ref{chm-inverse-phi-scaling}): $E(k) \sim E_{\phi}(k) \sim k^{-11/3}$.

For $k_i \ll k \ll 1$, a forward cascade of the free energy occurs, which takes the form of the cascade of enstrophy $Z$ and the passive cascade of $W_{\tilde{g}}$ (the invariant due to the zero-density part of $g$, equation \ref{g-ansatz}).  The spectrum of the potential fluctuations is $E_{\phi}(k) \sim k^{-5}$ (equation \ref{phi-forward-scaling-chm}).  To determine the scaling of $W_{g}(k)$, we must assume something about the strength at which $W_{\tilde{g}}$ is driven.  Let us suppose that the forcing drives $W_{\tilde{g}}$ weakly enough, such that we may neglect $W_{\tilde{g}}(k)$ in equation \ref{Wg-CHM-spec-reln}.  Then, for $k \ll \tau^{1/2}$, we have $W_{g}(k) \approx \tau E_{}(k) \approx \tau^2 E_{\varphi}(k)$ and for $\tau^{1/2} \ll k \ll 1$ we have $W_{g}(k) \approx \ZCHM(k) \approx k^4 E_{\varphi}(k)$.  The forward cascades of $\ZCHM$ and $W_{\tilde{g}}$ will combine around Larmor radius scale $k \sim 1$, to drive the cascade of $W_g$ in the nonlinear phase-mixing regime $k \gg 1$, giving $W_g(k) \sim k^{-4/3}$ and $E(k) \sim E_{\phi}(k) \sim k^{-10/3}$ (see section \ref{nl-mix-range-phenom-sec}, equations \ref{g-ell-scaling-phenom} and \ref{phi-scaling-phenom}).  These scalings are summarised in figure \ref{large-inject-fig}.

\subsubsection{Injection at small scales}

\begin{figure}
\begin{center}
\includegraphics[width=\columnwidth]{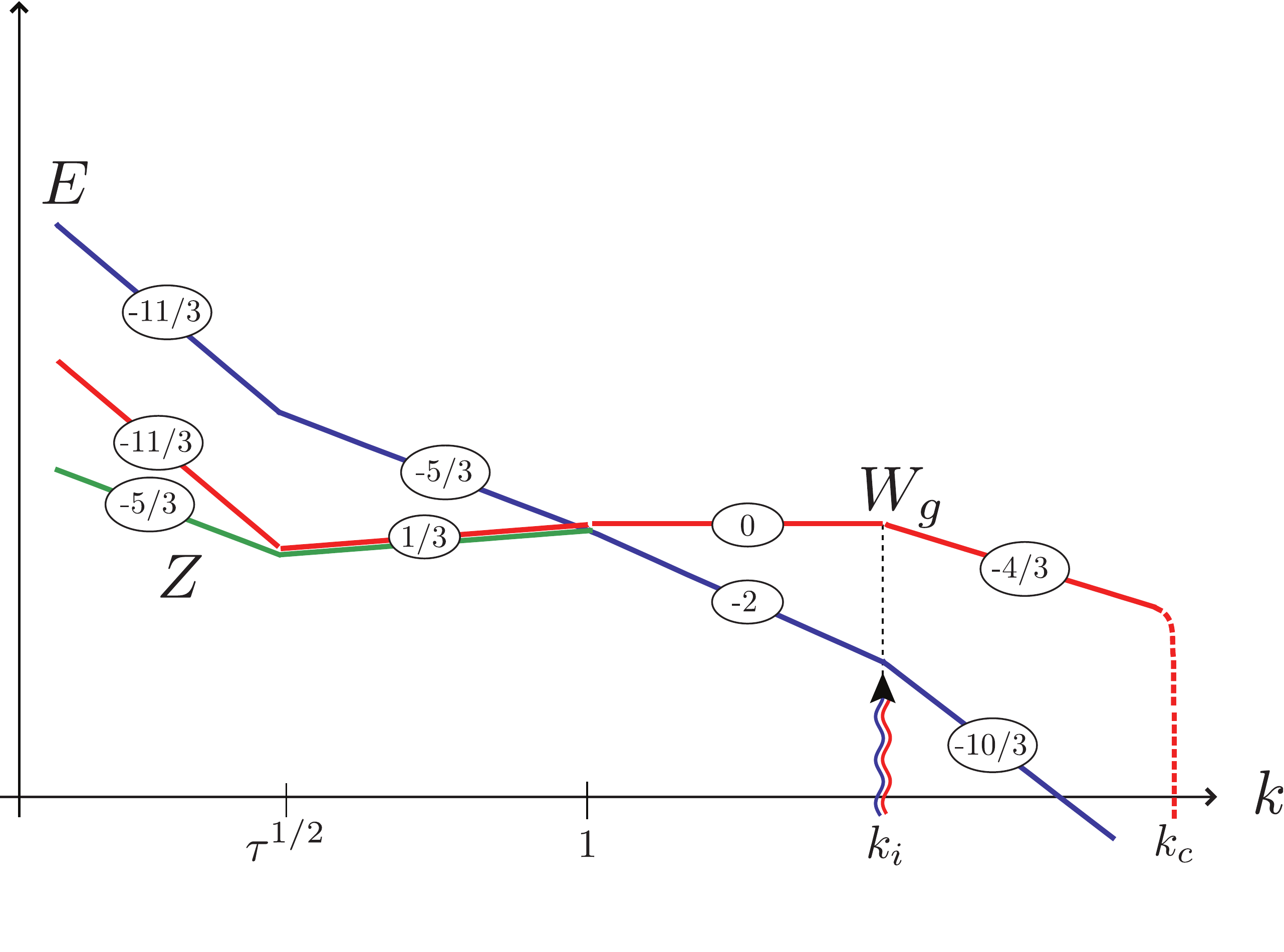}
\end{center}
\caption{Scenario 2: Injection at small scales.  The scaling indices for each sub-range are printed in bubbles over the plots of the invariants.  (Colour online.)}
\label{small-inject-fig}
\end{figure}

In figure \ref{small-inject-fig}, the injection is at small scales, in the nonlinear phase-mixing range, $k \gg 1$.  The free energy $W_{g}$ cascades forward to the collisional cutoff scale, $k_c$ (equation \ref{dissip-scale-eqn}), whereas the electrostatic invariant $E_{}$ cascades inversely.  For $1 \ll k \ll k_i$, the scaling of $E_{}(k)$ and $W_g(k)$ are taken directly from the phenomenology (section \ref{nl-mix-range-phenom-sec}, equations \ref{phi-scaling-phenom-inverse} and \ref{g-ell-scaling-phenom-inverse}): $E(k) \sim k^{-2}$ and $W_g(k) \sim k^0$.

Again, the identity of $W_{g}$ changes in the long wavelength limit, $k \ll 1$.  All fluctuations at these scales are due to the inverse cascade of $E_{}$, so $W_g(k)$ is determined from $E_{\phi}(k)$ by via equation \ref{Wg-CHM-spec-reln}.  The inverse cascade is characterised by the scaling $E_{\phi}(k) \sim k^{-11/3}$ (equation \ref{chm-inverse-phi-scaling}).  We argue that there is no transfer of $W_{\tilde{g}}$ from small scales because $\tilde{g}$ is passive.  Thus $W_{\tilde{g}}(k)$ is absent in equation \ref{Wg-CHM-spec-reln}.  In the kinetic limit of the CHM equation, $\tau \ll k^2 \ll 1$, we have $W_{g} \approx \ZCHM(k) \approx k^4 E_{\varphi}(k) \sim k^{1/3}$ and $E(k) \approx k^2 E_{\phi}(k) \sim k^{-5/3}$.  In the potential limit $k^2 \ll \tau$, we have $W_{g} \approx \tau E(k) \approx \tau^2 E_{\varphi}(k) \sim k^{-11/3}$, $E(k) \approx \tau E_{\phi}(k) \sim k^{-11/3}$ and $\ZCHM(k) \approx \tau k^2 E_{\phi}(k) \sim k^{-5/3}$.  These scalings are summarised in figure \ref{small-inject-fig}.

As we have shown in this section, much about the gyrokinetic turbulence can be understood by simple heuristic arguments.  Yet it is important to put these results onto firm foundation by making more precise, quantitative derivations.  We also would like to develop a more detailed theory of the phase-space structure of turbulent fluctuations.  This is done in sections \ref{statistical-sec}, \ref{formalism-sec} and \ref{spectral-theory-sec}.

\section[Exact laws]{Statistical theory}\label{statistical-sec}

In this section, we will analyse the gyrokinetic system, equations \ref{gyro-g} and \ref{qn-g}, using familiar statistical methods from fluid turbulence theory.  In sections \ref{symmetry-sec} and \ref{ensemble-sec}, we discuss symmetries of gyrokinetic system and the related statistical assumptions.  Sections \ref{third-order-result-W-sec} and \ref{exact-inverse-sec} will be devoted to deriving exact statistical scaling laws in analogy to Kolmogorov's `four-fifths' law (or the `three-halves' law for two-dimensional NS turbulence).  There will be two scaling laws: one for the forward cascade of the free energy $W_g$ and one for the inverse cascade of the electrostatic invariant $E_{}$.  Note that the exact laws constitute a precise statement of the constancy of nonlinear flux, which was used in the phenomenology of section \ref{nl-mix-range-phenom-sec}.

\subsection{Symmetries of the gyrokinetic system}\label{symmetry-sec}

Statistical arguments in turbulence theory are anchored to the underlying symmetries of the dynamical equations of interest \cite[see][]{frisch}.  Although specific boundary conditions, initial conditions and forcing mechanisms break these symmetries at large scales, a guiding principle of turbulence theory is that at sufficiently small scales, far from boundaries, a fully developed state will emerge where the symmetries of the dynamical equations are restored in a statistical sense.  We will appeal to this principle in the arguments of this paper.

As in fluid turbulence, one uses the translational invariance in time and position space for arguing statistical homogeneity and stationarity respectively; likewise, statistical isotropy originates from the rotation and reflection symmetries.  For gyrokinetics, the scaling symmetry is also particularly interesting.  In the collisionless limit, and for $k \gg 1$, the gyrokinetic system is invariant under the transformation

\begin{equation}
(g,\; \phi,\; {\bf r},\; v) \rightarrow (g,\; \mu^2\phi,\; \mu {\bf r},\; \mu v), \label{scaling-invariance}
\end{equation}

\noindent where $\mu$ is a scaling factor.  We will argue that this symmetry implies a dual-scaling in velocity and position space (see section \ref{dual-self-sim-sec}).  \footnote{This symmetry also provides a formal basis for the phenomenological argument $\ell_v \sim \ell$.  See section \ref{nl-mix-range-phenom-sec}, figure \ref{v-correl-fig}.}

\subsection{The ensemble average}\label{ensemble-sec}

We denote the ensemble average with an over-bar and define the correlation function of $g$ between two points in phase space, $({\bf R},\; v)$ and $({\bf R}^{\prime},\; v^{\prime})$, as

\begin{equation}
G =  \ensbl{ g({\bf R}, v_{}) g({\bf R}^{\prime}, v_{}^{\prime}) } = \ensbl{gg^{\prime}}.
\end{equation}

\noindent It is also useful to define the second order structure function:

\begin{equation}
\Stwo = \ensbl{\delta g^2},
\end{equation}

\noindent where $\delta g = g({\bf R}^{\prime}, v_{}^{\prime}) - g({\bf R}, v_{})$.  Homogeneity in position space (gyro-centre space) implies that $G$ and $\Stwo $ are functions only of the increment $\ellvec = {\bf R}^{\prime} - {\bf R}$.  In contrast, velocity space is fundamentally inhomogeneous, as there is no translational invariance in it.\footnote{Velocity dependence in the Maxwellian and the gyro-average both break translational invariance.}  However, we expect approximate ``local homogeneity'' in velocity space for sufficiently small velocity increments $\ell_v = v^{\prime} - v$.  Specifically, the two-point statistical quantities, such as structure and correlation functions, should vary strongly with the increment $\ell_v$ but weakly (smoothly) with the centred velocity $\bar{v} = (v^{\prime} + v)/2$.

\subsection{Third-order Kolmogorov relation for free energy}\label{third-order-result-W-sec}

We first derive the budget equation at scales $\ell$ and $\ell_v$.  We take the gyrokinetic equation \ref{gyro-g} for $g$ and $g^{\prime}$ and multiply by $g^{\prime}$ and $g$, respectively, then add the two resulting equations to obtain

\begin{equation}
  \label{budget_real2}
\frac{\partial G}{\partial t} - \frac{1}{2}\bnabla_{\ell_{}}\cdot\Sthree = \ensbl{g\CollisionOpP{h^{\prime}} + g^{\prime}\CollisionOp{h}} + \ensbl{g\Source^{\prime}} + \ensbl{g^{\prime}\Source},
\end{equation}

\noindent where the third-order structure function is defined 

\begin{equation}
\Sthree = \ensbl{\delta{\bf v}_E\delta g^2}\label{s3-def}
\end{equation}

\noindent and $\delta{\bf v}_E = {\bf v}_E^{\prime} - {\bf v}_E$.  To get equation \ref{budget_real2}, we have used incompressibility of ${\bf v}_E$ ($\bnabla\cdot{\bf v}_E = 0$) and statistical homogeneity to manipulate the nonlinear flux to obtain $-\ensbl{g^{\prime}{\bf v}_E\cdot\bnabla g} + \ensbl{g{\bf v}_E^{\prime}\cdot\bnabla^{\prime} g^{\prime}} = \bnabla_{\ell_{}}\cdot\ensbl{\delta{\bf v}_E\delta g^2}/2$.

Now if we take $\ellvec_{} = 0$, $\ell_v = 0$, the nonlinear term in equation \ref{budget_real2} is zero by homogeneity.  The result is the kinetic `global budget' equation:

\begin{equation}
\frac{\partial\ensbl{g^2}}{\partial t} = 2\ensbl{g\CollisionOp{h}} + 2\ensbl{g\Source}.
\label{budget_global_t}
\end{equation}

\noindent Assuming stationarity, the left-hand side of the equation is zero and we have the steady state balance between forcing and dissipation

\begin{equation}
\ensbl{g\Source} = -\ensbl{g\CollisionOp{h}} = \varepsilonW(\bar{v_{}}),
\label{budget_global}
\end{equation}

\noindent where we have defined the kinetic free energy injection rate $\varepsilonW(\bar{v_{}})$.  

We now assume that the forcing is restricted to a narrow range around an injection scale which we denote \ellF.  For $\ell \ll \ellF$, $\ell_v \ll \ellF$ and $\ell_v \ll 1$, the forcing term in the budget equation \ref{budget_real2} reduces to

\begin{equation}
\ensbl{g\Source^{\prime}} + \ensbl{g^{\prime}\Source} \approx 2\ensbl{g\Source} = 2\varepsilonW(\bar{v_{}}).
\end{equation}

\noindent Now we take the limit of small collisionality: the collision operator is proportional to $\nu$, the collision frequency, which we take to be small so that, for fixed $\ell$ and $\ell_v$, the collisional term is negligible in the budget equation \ref{budget_real2}.  \footnote{What is implicit to this assumption is that a collisional cutoff scale $\ell_c$ exists, such that for $\ell, \ell_v \gg \ell_c$, the collisional term may be neglected --- the existence of this scale can either be demonstrated {\em a posteriori} or assumed from the arguments of the phenomenology section (see equation \ref{dissip-scale-eqn}).} Thus, by assuming small collisionality, statistical stationarity, $\ell_v \ll 1$ and $\ell \ll \ellF$ we have from equation \ref{budget_real2}

\begin{equation}
\bnabla_{\ell_{}}\cdot\Sthree = -4\varepsilonW(\bar{v_{}}).
\label{four-fifths}
\end{equation}

\noindent Note that although homogeneity and stationarity have been assumed, we have not yet assumed isotropy.  Assuming it now, we may write $\Sthree = \hat{\ellvec} \SthreePl(\ell)$, where $\hat{\ellvec} = \ellvec/\ell$.  Then equation \ref{four-fifths} reduces to scalar form:

\begin{equation}
\frac{1}{\ell}\frac{\partial}{\partial \ell}\left(\ell \;\SthreePl\right) = -4\varepsilonW(\bar{v_{}}).
\label{four-fifths-scalar-eqn}
\end{equation}

\noindent The solution to equation \ref{four-fifths-scalar-eqn} which satisfies $\SthreePl (\ell = 0) = 0$ is

\begin{equation}
\SthreePl = -2\varepsilonW(\bar{v_{}})\ell.
\label{four-fifths-iso}
\end{equation}

\subsection{Third-order Kolmogorov relation for the electrostatic invariant}\label{exact-inverse-sec}

To derive the scale-by-scale budget equation for the electrostatic invariant $E_{}$ we proceed similarly to the analysis of section \ref{third-order-result-W-sec}.  However, in this case, we are dealing with functions of real-space variable ${\bf r}$ and its increment $\ellvec_{r} = {\bf r}^{\prime} - {\bf r}$.  Also, velocity will appear only as a variable of integration, so it will simplify our notation to refer to a single variable $v$ in what follows --- \ie ${\bf R} = {\bf r} - \rhovec(v)$, ${\bf R}^{\prime} = {\bf r}^{\prime} - \rhovec(v)$ and $g^{\prime} = g({\bf R}^{\prime}, v)$, \etc , from which it follows that $\ellvec_{} = \ellvec_{r}$ and we will refer to a single increment in position space, $\ellvec$.

We take the gyrokinetic equation \ref{gyro-g} for $g$ and $g^{\prime}$ and multiply by $\phi^{\prime}$ and $\phi$ respectively, then ensemble-average and integrate over velocity $v_{}$.  The result is

\begin{multline}
(1 + \tau - \Gamma_0)\frac{\partial \Phi}{\partial t} - \bnabla_{\ell_{}}\cdot\SthreeE = \\ 2\pi \; \ensbl{\phi^{\prime}\int v dv\angleavg{\CollisionOp{h}}} + 2\pi \; \ensbl{\phi \int v dv \angleavgP{\CollisionOpP{h^{\prime}}}} + \ensbl{\phi\SourceE^{\prime}} + \ensbl{\phi^{\prime}\SourceE},
\label{budget-E}
\end{multline}

\noindent where we have defined the correlation function $\Phi = \ensbl{\phi \phi^{\prime}}$ and introduced the third-order structure function

\begin{equation}
\SthreeE(\ell) =  2\pi \ensbl{\int v dv \; \delta\gyroavg{\phi} \delta {\bf v}_E\delta g },\label{sthree-E-def}
\end{equation}

\noindent where $\delta \gyroavg{\phi} = \gyroavgP{\phi^{\prime}} - \gyroavg{\phi}$.  Note that by homogeneity, we have $\Gamma_0 \Phi = \Gamma_0^{\prime} \Phi = \Gamma_0^{(\ell_{})}\Phi$, where $\Gamma_0^{(\ell_{})}$ operates in $\ell_{}$-space.  Lastly, we have defined the electrostatic forcing $\SourceE = 2\pi \int vdv \angleavg{\Source}$.  

We can express the first term on the right side of equation \ref{budget-E} in terms of the rate of change of the invariant $E_{}$ and structure functions involving $\phi$.  Using homogeneity, we find the following relation

\begin{equation}
(1 + \tau - \Gamma_0)\Phi = 2E - \frac{1}{2}\left(\ensbl{\delta \phi^2} - \ensbl{\delta \phi \delta(\Gamma_0\phi)}\right),\label{dedt-eqn}
\end{equation}

\noindent where $\delta(\Gamma_0 \phi) = \Gamma_0^{\prime}\phi^{\prime} - \Gamma_0\phi$.  In the theory of fluid turbulence, it is standard at this stage to neglect the time derivatives of structure functions in the stationary limit \cite[\eg][]{kolmogorov41c}, assuming essentially that a self-similar regime is attained.  We follow this practice and substitute $(1 + \tau - \Gamma_0)\partial_t\Phi \approx 2\partial_tE_{}$ into equation \ref{budget-E}.  Taking the limit of small collisionality, equation \ref{budget-E} becomes

\begin{equation}
\frac{d E}{d t} = \frac{1}{2}\bnabla_{\ell_{}}\cdot\SthreeE + \frac{1}{2}\ensbl{\phi\SourceE^{\prime}} + \frac{1}{2}\ensbl{\phi^{\prime}\SourceE}.\label{budget-E-stationary}
\end{equation}

\noindent For $\ell = 0$, we have the global balance of $E_{}$, which describes the continual accumulation due to forcing:

\begin{equation}
\frac{d E}{d t} = \ensbl{\phi\SourceE}\label{global-budget-E} \equiv \varepsilonE,
\end{equation}

\noindent where $\varepsilonE$ is the injection rate of the electrostatic invariant $E_{}$.  Recall that the forcing is assumed to be restricted to a scale $\ellF$ as explained in the section \ref{third-order-result-W-sec}.  For the inverse cascade, we are considering an inertial range such that $\ell \gg \ellF$.  In this limit, the forcing term in equation \ref{budget-E-stationary} is zero and we find the following exact (anisotropic) result:

\begin{equation}
\bnabla_{\ell_{}}\cdot \SthreeE = 2\varepsilonE.
\label{four-fifths-E}
\end{equation}

\noindent Assuming isotropy, we have $\SthreeE = \hat{\ellvec}\SthreeEPl(\ell)$ and so equation \ref{four-fifths-E} gives

\begin{equation}
\SthreeEPl = \varepsilonE\ell.
\label{four-fifths-E-iso}
\end{equation}

\noindent Note that in the CHM/Euler limit ($\ell \gg 1$), equation \ref{four-fifths-E} may be manipulated to coincide with the result of \cite{boffetta} for CHM turbulence and the result of \cite{bernard} for two-dimensional NS turbulence: $\bnabla_{\ell_{}}\cdot\ensbl{\delta {\bf u} |\delta {\bf u}|^2} = 4\varepsilonE$, where ${\bf u} \equiv {\bf v}_{E0} = {\bf {\hat z}}\times\bnabla\varphi$.  

The two exact laws derived in this section, equations \ref{four-fifths-iso} and equation \ref{four-fifths-E-iso}, hold for arbitrary spatial scales, \ie, they hold above, below and at the Larmor radius scale.  (Note, however, that for equation \ref{four-fifths-iso}, we have assumed that the velocity scale satisfies $\ell_v \ll 1$.)  In the sections that follow, we will consider the limit $\ell_{}, \ell_v \ll 1$.   This is the essentially kinetic limit (see section \ref{nl-mix-range-phenom-sec} for phenomenology) where nonlinear phase-mixing is dominant and the turbulent cascade involves both velocity and position space.

\section[Spectral formalism]{Spectral formalism}\label{formalism-sec}

In this section we introduce a way of characterising velocity scales using a zeroth-order Hankel transform.  This produces an elegant spectral representation of gyrokinetics, where the electrostatic potential and electrostatic invariant are compactly expressed in terms of the Hankel transform of the distribution function.  In section \ref{spectral-theory-sec} we will use this formalism to develop a spectral scaling theory for the inertial ranges in phase-space.

\subsection{Hankel transform}

The Hankel transform of a function $g(v)$ is

\begin{equation}
\hat{g}(p) = \int_0^{\infty} vdvJ_0(pv)g(v).
\end{equation}

\noindent The Hankel transform is its own inverse.  This is easily proved using the orthogonality of the zeroth-order Bessel functions:

\begin{equation}
\int_0^{\infty} J_0(p_1 x) J_0(p_2 x) x dx = \delta(p_1 - p_2)/p_1.\label{orth-bessel}
\end{equation}

\noindent Using this identity, the Plancherel theorem for two functions $f$ and $g$ is

\begin{equation}
\int_0^{\infty} v dv f(v) g(v) = \int_0^{\infty} p dp \hat{f}(p)\hat{g}(p).\label{gen-parseval}
\end{equation}

The following integral involving three Bessel functions will be useful in deriving the mode coupling due to the nonlinearity:

\begin{equation}
\int_0^{\infty} v dv J_0(p_1 v)J_0(p_2 v)J_0(p_3 v) = K(p_1,p_2, p_3), \label{three-bes-int}
\end{equation}

\noindent where if $p_1$, $p_2$ and $p_3$ do not form the sides of a triangle, then $K = 0$; if they do, then $K = 1/(2\pi\Delta)$, where $\Delta$ is the area of that triangle.  

\subsection{Spectral gyrokinetic equation}

With the help of the above identities, we may now rewrite the gyrokinetic system as a single equation in Hankel--Fourier space.  The Hankel--Fourier transform of the distribution function $g({\bf R}, v_{})$ is

\begin{equation}
\hat{g}({\bf k}, p) \equiv \frac{1}{2\pi}\int_{\field{R}}d^2{\bf R}\int_0^{\infty}v_{}dv_{}J_0(pv_{})\e^{-\i{\bf k}\cdot{\bf R}}g({\bf R},v_{}).\label{hank-four-def}
\end{equation}

\noindent Quasi-neutrality, equation \ref{qn-g-k-1}, can now be expressed simply as

\begin{equation}
\hat{\varphi}({\bf k}) = \beta({\bf k})\hat{g}({\bf k}, k).\label{qn-k-p}
\end{equation}

\noindent Neglecting collisions and applying the Hankel--Fourier transform to the gyrokinetic equation \ref{gyro-g}, we find the spectral form of the gyrokinetic system

\begin{equation}
\frac{\partial \hat{g}({\bf k}, p)}{\partial t} =  \frac{1}{2\pi}\int d^2{\bf k}^{\prime} \beta(k^{\prime})\; {\bf k} \cdot({\bf {\hat z}}\times{\bf k}^{\prime}) \int qdq \; K(k^{\prime}, p, q) \; \hat{g}({\bf k}^{\prime}, k^{\prime}) \hat{g}({\bf k}-{\bf k}^{\prime}, q),
\label{coupling}
\end{equation}

\noindent where $\beta$ is defined in equation \ref{beta-def-eqn} and $K$ is given in equation \ref{three-bes-int}.

\subsection{Free energy spectrum}

\noindent Let us apply this formalism to define the spectral density of free energy.  First, we use homogeneity and isotropy to express the conserved quantity $\ensbl{g^2}$ in terms of the correlation function $G(\ell_{},v_{}^{\prime}, v_{})$:

\begin{eqnarray}
\ensbl{g^2}(v)& = \int k dkpdp\ell_{} d\ell_{} v_{}^{\prime}dv_{}^{\prime}J_0(k\ell_{})J_0(pv_{}^{\prime})J_0(pv_{})G(\ell_{} ,v_{}^{\prime},v_{}),
\label{spectral-manip}
\end{eqnarray}

\noindent where we have used equation \ref{orth-bessel} and the identity $\int d^2{\bf k} \e^{\i {\bf k}\cdot{\bf r}} = (2\pi)^2\delta ({\bf r})$.  Note that isotropy has been used to reduce the Fourier transform in vector position increment $\ellvec$ to a Hankel transform in scalar increment $\ell$.  For simplicity, we consider generalised free energy corresponding to $\kappa = 1$ (see equation \ref{free-energy-a}):

\begin{equation}
W_{g1} = 2\pi\int v_{}dv_{}\frac{\ensbl{g^2}}{2}.
\end{equation}

\noindent However, we stress that the spectral scaling results derived below will hold for general $\kappa(v)$ as long as this function has weak dependence on velocity (\ie for $\kappa(v)$ with $\ell_v \sim 1$).  Thus, scaling results will be reported in terms of $W_g(k,p)$, although $W_{g1}$ will be used during intermediate steps, for the sake of consistent notation.  

From equation \ref{spectral-manip} it follows that 

\begin{equation}
W_{g1} = \int dkdp\; W_{g1}( k,p),
\end{equation}
 
\noindent where

\begin{equation}
W_{g1}(k,p) = \pi pk\int \ell_{}d\ell_{}v_{}dv_{}v_{}^{\prime}dv_{}^{\prime}J_0(k\ell_{})J_0(pv_{}^{\prime})J_0(pv_{})G(\ell_{},v_{}^{\prime},v_{}).
\label{sg-def}
\end{equation}

\noindent $W_{g1}(k,p)$ can be re-expressed in more convenient velocity variables: the increment variable $\ell_v = v_{}^{\prime} - v_{}$ and the centred velocity $\bar{v} = (v_{} + v_{}^{\prime})/2$.  Then equation \ref{sg-def} becomes

\begin{equation}
W_{g1}(k,p) = \pi pk\int \ell_{} d\ell_{}(\bar{v}^2 -  \ell_v^2/4)d\ell_v d\bar{v} J_0(k\ell_{})Q(p,\ell_v,\bar{v})G(\ell_{}, \ell_v, \bar{v}),
\label{sg-spectrum-def}
\end{equation}

\noindent where $Q(p, \ell_v, \bar{v}) = J_0[p(2\bar{v} + \ell_v )/2]J_0[p(2\bar{v} - \ell_v)/2]$.

Note that equation \ref{sg-def} is not invertible, unlike the case of the Wiener-Khinchin formula in fluid turbulence --- \ie, the correlation function $G(\ell_{},v_{}^{\prime},v_{})$ cannot be recovered from $W_g(k,p)$.  Because of the lack of homogeneity in $v_{}$, we have integrated over $v_{}$ and $v_{}^{\prime}$ to define scales in velocity space via a single velocity-space ``wavenumber'' $p$. 

The spectral density can also be expressed in terms of the Hankel--Fourier transform of the distribution function (using equations \ref{gen-parseval} and \ref{hank-four-def}):  

\begin{eqnarray}
W_{g1}(k, p)  = \pi\;pk\;\ensbl{\hat{g}(k,p)^2}.\label{w-spec-def-eqn}
\end{eqnarray}

\subsection{Spectrum of the electrostatic invariant}

Using equations \ref{E-def} and \ref{qn-k-p}, the spectral density of the electrostatic invariant can now be expressed

\begin{equation}
E(k) = \frac{\pi k}{\beta(k)}\ensbl{|\hat{\varphi}({\bf k})|^2} = \frac{\beta(k)}{k}W_{g1}(k, k),\label{E-W-spec-reln}
\end{equation}

\noindent which also implies that $E_{\varphi}(k) = \beta^2(k) W_{g1}(k,k)/(2\pi k)$.  We can also write $\StwoE = \ensbl{|\delta{\bf v}_E|^2}$ in terms of the spectral free energy density $W_{g1}(k,p)$.  Using equation \ref{qn-k-p}, we find

\begin{equation}
\StwoE  = \int dk H(k, \ell_{}, v_{}, v_{}^{\prime}) W_{g1}(k, k),
\label{EB-struc-fcn}
\end{equation}

\noindent where $H(k, \ell_{}, v_{}, v_{}^{\prime}) = -k \beta^2(k)/\pi\left[J_0^2(k v_{}^{\prime}) + J_0^2(k v_{}) - 2 J_0(k v_{}^{\prime})J_0(k v_{})J_0(k\ell_{})\right]$.

\section{Spectral scaling theory}\label{spectral-theory-sec}

In section \ref{formalism-sec}, we did not make any assumption about the size of $k$ or $p$ and the equations presented were exact.  In this section, we will derive some approximate spectral scaling results at the sub-Larmor scales, $k \gg 1$ and $p \gg 1$.  In section \ref{dual-self-sim-sec} we introduce a similarity hypothesis for phase space that will allow us to infer spectral scalings from the exact laws of section \ref{statistical-sec}.  In section \ref{spectral-scaling-sec}, we derive general relationships between the scaling of the structure function, $\Stwo$, and spectral densities $W_g(k)$ and $E(k)$.  Then, in section \ref{spectra-forward-sec}, we derive specific scaling laws for the forward-cascade range and, in section \ref{inverse-cascade-sec}, for the inverse-cascade range.

\subsection[Self-similarity]{Self-similarity hypothesis and scaling in velocity space}

\label{dual-self-sim-sec}

In the theory of fluid turbulence, Kolmolgorov's four-fifths law can be used to infer the scaling index for the velocity field under the assumption of self-similarity.  This allows derivation of the (approximate) energy spectrum directly from the exact result for the third-order structure function.  We will follow an analogous approach for gyrokinetics in the following sections.  In our case, however, we would like to determine scaling with respect to the increment $\ell_v$ in addition to $\ell$ (and corresponding phase-space spectral scaling laws).  To accomplish this, we supplement the exact laws, equations \ref{four-fifths-iso} and \ref{four-fifths-E-iso}, with the following {\it dual self-similarity hypothesis}: given fixed scales $\ell$ and $\ell_v$, such that $\ell_v, \ell \ll 1$, the increment $\delta g$ obeys

\begin{equation}
\delta g(\lambda \ell_{}, \lambda \ell_{v}) \doteq \lambda^{h_g}\delta g(\ell_{}, \ell_{v}).
\label{dual-hypothesis}
\end{equation}

\noindent The notation $\doteq$ is meant to echo the ``equality in law'' of \cite{frisch}.  This means that $\delta g$ exhibits this scaling statistically, \ie whenever it appears inside an ensemble average.  For instance, it implies $\Stwo(\lambda \ell, \lambda \ell_v)/\Stwo(\ell, \ell_v) = \lambda^{2h_g}$.  We argue that equation \ref{dual-hypothesis} is true on symmetry grounds.  In particular, since the collisionless gyrokinetic system is invariant to simultaneous scaling of ${\bf R}$ and $v$ by the same factor (this is the symmetry \ref{scaling-invariance}), the statistics of the stationary state should obey this symmetry.\footnote{Note that, although $\delta g$ must depend on the centred velocity $\bar{v}$, its dependence is weak, and thus we argue that the scaling symmetry is only in $\ell$ and $\ell_v$.  The sensitive dependence in $\ell_v$ is due to the fact that the fluctuations must decorrelate in velocity space on the scale $\ell_v \sim \ell$, as argued by \cite{schek-ppcf} and reiterated in section \ref{nl-mix-range-phenom-sec} with illustration in figure \ref{v-correl-fig}.  One can think of this as ``local homogeneity,'' though it is clearly different from homogeneity in the sense of neutral fluid turbulence.}

As we prove below (see equations \ref{EB-struc-fcn-approx} and \ref{he-nu}), the dual self-similarity hypothesis for $\delta g$ implies self-similarity for the gyro-averaged ${\bf E} \times {\bf B}$ velocity difference $\delta {\bf v}_E$:

\begin{equation}
\delta {\bf v}_E(\lambda \ell, \lambda \ell_v) \doteq \lambda^{\hE}\delta {\bf v}_E(\ell, \ell_v).
\label{stwoe-scaling}
\end{equation}

\subsection[Spectral scaling in phase-space]{Spectral scaling in the inertial ranges}\label{spectral-scaling-sec}

Consider equation \ref{sg-spectrum-def}.  To begin, we take $p\bar{v} \gg 1$ so that we can use the large-argument approximation of the Bessel functions in $Q(p, \ell_v, \bar{v})$.  Also, we assume that the correlation function $G(\ell, \ell_v, \bar{v})$ is peaked at small $\ell_v$, so the integral is dominated by $\ell_v \ll \bar{v}$.  Since the integral is dominated by small $\ell_v$, the bounds of the integration over $\ell_v$ are unimportant.  In this limit,

\begin{equation}
Q(p, \ell_v, \bar{v}) \approx \frac{1}{4\pi p \bar{v}}\left[\sin(2 p \bar{v}) + \cos(p \ell_v)\right].
\end{equation}

\noindent Substituting this expression into equation \ref{sg-spectrum-def}, we see that the term proportional to $\sin(2 p \bar{v})$ oscillates rapidly in $\bar{v}$ --- while other quantities in the integrand depend only weakly on $\bar{v}$ --- so this term is negligible.\footnote{This statement can be confirmed {\em a posteriori} by considering the scaling results given by equation \ref{scaling-indices-solutions}.  For instance, if we consider $\ell_{} = 0$, we then have $\Stwo \sim |\ell_v|^{1/3}$ and the term we have neglected is indeed smaller by a factor of $p^{-2}$.}  Then, using $G = [\ensbl{g^2} + \ensbl{(g^{\prime})^2} - \Stwo]/2$, we can express $W_g(k,p)$ in terms of $\Stwo = \ensbl{\delta g^2}$, for non-zero $k$ or $p$:

\begin{equation}
W_{g1}(k,p) \approx -\frac{k}{8}  \int \ell_{} d\ell_{}\; \bar{v} d\bar{v} \; d\ell_v J_0(k\ell_{})\cos(p\ell_v)\;\Stwo(\ell, \ell_v, \bar{v}).
\label{sg-def-approx}
\end{equation}

\noindent Then, dual self-similarity of $\delta g$ (equation \ref{dual-hypothesis}) implies that $W_{g1}(k, p)$, and, therefore, $W_g(k, p)$, must satisfy the scaling

\begin{equation}
W_g(\lambda k, \lambda p) = \lambda^{\nu}W_g(k, p),
\label{sg-dual-scaling}
\end{equation}

\noindent with

\begin{equation}
\nu = -2 - 2h_g.
\label{nu-hg}
\end{equation}

\noindent We will now determine the relationship between the scaling indices $\nu$ and $\hE$ (defined in equation \ref{stwoe-scaling}).  Expanding equation \ref{EB-struc-fcn} in the limit $k\bar{v} \gg 1$ and $\ell_v \ll \bar{v}$, as done for $Q$ above, we get

\begin{equation}
\StwoE  \approx \frac{2}{(1 + \tau)^2\bar{v}}\int dk \left[1 - \cos(k\ell_v)J_0(k\ell_{})\right] W_{g1}(k, k).
\label{EB-struc-fcn-approx}
\end{equation}

\noindent Using equations \ref{EB-struc-fcn-approx} and \ref{sg-dual-scaling}, we find:

\begin{equation}
2\hE = -1 - \nu.
\label{he-nu}
\end{equation}

\subsection{Spectral scaling in the forward-cascade range}\label{spectra-forward-sec}

As a consequence of the dual self-similarity hypothesis and the exact third-order law, equation \ref{four-fifths-iso}, we have the following scaling law for $\SthreePl$:

\begin{equation}
\SthreePl(\lambda \ell_{}, \lambda \ell_v) = \lambda \SthreePl(\ell_{}, \ell_v)
\end{equation}

\noindent Thus, the scaling indices for $\delta g$ and $\delta {\bf v}_E$ must satisfy the following relation in the forward-cascade range:

\begin{equation}
2h_g + \hE = 1.
\label{hg-he}
\end{equation}

\noindent Combining this with equations \ref{nu-hg} and \ref{he-nu}, we have

\begin{subequations}\label{scaling-indices-solutions}
\begin{align}
&h_g = 1/6,\label{hg-solution} \\
&\hE = 2/3, \\
&\nu = -7/3.
\end{align}
\end{subequations}

\noindent And, therefore, taking $p = k$, we have the power law

\begin{equation}
W_g(k, k) \propto k^{-7/3}.
\label{sg-kk-scaling}
\end{equation}

\noindent This implies (see equation \ref{E-W-spec-reln}) that in the large-$k$ limit,

\begin{equation}
E_{}(k) \approx (1 + \tau) E_{\phi}(k) = \frac{1}{k(1+\tau)} W_{g1}(k,k) \propto k^{-10/3}.
\end{equation}

\noindent This agrees with the $E_{\phi}(k)$ scaling derived phenomenologically in section \ref{nl-mix-range-phenom-sec} \cite[equation \ref{phi-scaling-phenom}; also, see][]{schek-ppcf, schekochihin}, and has been confirmed numerically by \cite{tatsuno-prl, tatsuno-pop}.  

\subsubsection{Phase-space spectrum $W_g(k, p)$}\label{two-dim-spectrum-sec}

We have found that the phase-space spectrum $W_g(k, p)$ obeys the scaling law given by equation \ref{sg-dual-scaling}, with $\nu = -7/3$.  To obtain this result, we have assumed $k \gg 1$ and $p \gg 1$, but have assumed nothing about the relative size of $k$ and $p$.  In the subsidiary limits $k \gg p$ and $k \ll p$, we can show that the spectrum $W_g(k, p)$ has power law scalings separately in $k$ and $p$.  These scalings, obtained in appendix \ref{spectra-deriv-sec}, are

\begin{equation}
W_g(k, p) \propto 
\begin{cases}
k^{-2}p^{-1/3}\mbox{,\hspace{2mm} for \hspace{2mm}} k \gg p \gg 1,\\
p^{-2}k^{-1/3}\mbox{,\hspace{2mm} for \hspace{2mm}} p \gg k \gg 1.
\end{cases}
\label{skp-limit-scaling}
\end{equation}

The one-dimensional spectra $W_g(k)$ and $W_g(p)$ (\ie, $W_g(k,p)$ integrated over $p$ and $k$, respectively) are also derived in appendix \ref{spectra-deriv-sec}.  We find the following power laws:

\begin{equation}
W_g(k) \propto k^{-4/3},
\label{sg-k-scaling}
\end{equation}

\noindent confirming the arguments of section \ref{nl-mix-range-phenom-sec} (equation \ref{g-ell-scaling-phenom}), and

\begin{equation}
W_g(p) \propto p^{-4/3}.
\label{sg-p-scaling}
\end{equation}

\noindent The power law for $W_g(k)$ agrees with the prediction by \cite{schek-ppcf, schekochihin}, while the power law for $W_g(p)$ is a new prediction.  Both scaling laws have been confirmed numerically by \cite{tatsuno-prl, tatsuno-pop}.  As a consistency check one may integrate the asymptotic forms from equation \ref{skp-limit-scaling} directly, to obtain roughly $W_g(p) \sim \int_0^p dk k^{-1/3}p^{-2} + \int_p^{\infty} dk k^{-2}p^{-1/3} \sim p^{-4/3}$ and similarly for $W_g(k)$.

\subsection{Spectral scaling in the inverse-cascade range}

\label{inverse-cascade-sec}

We now turn to the scaling of the spectra for the high-$k$ inverse-cascade range, $1 \ll k \ll k_i$.  As in section \ref{spectra-forward-sec}, we will combine the dual self-similarity hypothesis with an exact law; in this case, the exact law is equation \ref{four-fifths-E-iso}, the scaling law for $\SthreeEPl$.  However, there is now a complication arising from the velocity integration in the definition of $\SthreeE$, equation \ref{sthree-E-def}.  As we have argued in section \ref{nl-mix-range-phenom-sec}, integration of the distribution function effectively averages over the random velocity dependence and, thus, acts as a partial average, which we cannot assume commutes with the formal ensemble average.

From the random-walk argument of section \ref{nl-mix-range-phenom-sec}, we expect that the velocity integration of $\delta g$ should, roughly speaking, introduce a factor of $\ell^{1/2}$.  This `rule of thumb' has been validated in the forward-cascade range by the numerical results of \cite{tatsuno-prl, tatsuno-pop}; furthermore, it agrees with the results obtained using the Hankel transform treatment.  To formalise this, we introduce the scaling hypothesis

\begin{equation}
\int dv \; F(v) \; \delta g(\lambda\ell_{}, v) \doteq \lambda^{h_g + 1/2}\;\int dv \; F(v) \; \delta g(\ell_{}, v),\label{v-int-scale-hyp}
\end{equation}

\noindent where $F(v)$ is an arbitrary function, varying over scale $\ell_v \sim \ell$.  Also note that $\delta g(\ell, v) = g({\bf R}^{\prime}, v) - g({\bf R}, v)$ depends on a single velocity $v$.  

Applying the hypothesis \ref{v-int-scale-hyp} to the result for $\SthreeEPl$, equation \ref{four-fifths-E-iso}, we find the following relationship between the scaling indices

\begin{equation}
2\hE +  h_g = -1/2.\label{E-scaling-result}
\end{equation}

\noindent Combing this with equations \ref{nu-hg} and \ref{he-nu}, we get

%\subsubsection{\textcolor{blue}{Option 2: Do it without an extra hypothesis}}

%Our approach is to evaluate the velocity integration in equation \ref{sthree-E-def} by expanding  $\SthreeEPl$ in terms of third-order correlation functions and applying the Hankel spectral formalism.  Following the derivations of equations \ref{sg-def-approx} and \ref{EB-struc-fcn-approx}, we expand the Bessel functions in the large argument limit and assume that the correlation functions are peaked around $\ell_v \sim \ell \ll \bar{v}$.  (We omit the details as they are essentially a repetition of those found elsewhere in this paper.)  The result is an expression for $\SthreeEPl$ in terms of $\delta\phi$ and $\delta{\bf v}_{E0}$:

%\begin{equation}
%\SthreeEPl \approx \ensbl{\delta{\bf v}_{E0}\delta\phi^2}\;\psi_E(\ell),
%\end{equation}

%\noindent where $\psi_E(\ell)$ is a scaling function that satisfies $\psi_E(\lambda\ell)/\psi_E(\ell) = \lambda^{1/2}$.  By the usual arguments, it can be shown that $\delta\phi(\lambda \ell) \doteq \lambda^{-\nu/2} \delta{\phi}(\ell)$.  Then, using equation \ref{four-fifths-E-iso}, we find

%\subsubsection{\textcolor{blue}{End options}}

\begin{equation}
\nu = -1.\label{inverse-nu-result}
\end{equation}

\noindent Therefore

\begin{subequations}
\begin{align}
&W_g(k,k) \propto k^{-1},\label{inverse-W-k-scaling}\\
&E_{}(k) \propto k^{-2}.\label{inverse-E-k-scaling}
\end{align}
\end{subequations}

\noindent Equation \ref{inverse-E-k-scaling} agrees with the phenomenological result of section \ref{nl-mix-range-phenom-sec}, equation \ref{phi-scaling-phenom-inverse}.  Let us now determine the free-energy spectral scalings.  First note that by equations \ref{nu-hg} and \ref{inverse-nu-result}, we have

\begin{equation}
h_g = -1/2.
\end{equation}

\noindent Now, if the free energy spectrum is a power law in $k$-$p$ space as we have assumed in the forward cascade range (see equation \ref{skp-limit-scaling}), then we find

\begin{subequations}
\begin{align}
&W_g(k, p) \propto
\begin{cases}
k^{-2}p\mbox{,\hspace{2mm} for \hspace{2mm}} k \gg p\\
p^{-2}k\mbox{,\hspace{2mm} for \hspace{2mm}} k \ll p
\end{cases}\label{inverse-phase-spectra-laws-1}\\
& W_g(k) \propto k^{0},\label{inverse-phase-spectra-laws-2}\\
& W_g(p) \propto p^{0},\label{inverse-phase-spectra-laws-3}
\end{align}
\end{subequations}

\noindent where we have adapted the arguments of section \ref{two-dim-spectrum-sec} for the inverse-cascade result $h_g = -1/2$, expanding equation \ref{sg-def-approx} in the limits $\ell_v \ll \ell$ and $\ell_v \gg \ell$.  Equation \ref{inverse-phase-spectra-laws-2} agrees with the phenomenological result of section \ref{nl-mix-range-phenom-sec}, equation \ref{g-ell-scaling-phenom-inverse}.

\section{Conclusion}

\subsection{Summary}

In this paper, we have considered forced two-dimensional gyrokinetics as a simple paradigm for magnetised plasma turbulence.  We have explored how the nonlinear phase-mixing mechanism gives rise to a phase-space cascade of free energy to scales much smaller than the Larmor radius.  We have investigated the relationship between the inertial-range cascades in fluid theories (\ie the CHM and two-dimensional Euler equations) and gyrokinetics, and found a simple relationship between the fluid invariants and the kinetic invariants, given by equation \ref{invar-reln}.  The gyrokinetic free energy invariant $W_g$ (equation \ref{gen-free-energy-0}) is cascaded to fine scales and is ultimately dissipated by collisions.  We predict that, concurrently, the gyrokinetic electrostatic invariant $E_{}$ (equation \ref{E-def}) cascades inversely from very fine scales (\eg, the electron Larmor radius) to long wavelengths --- possibly even to wavelengths of the CHM/Euler limits, much larger than the ion Larmor radius.

In addition to a phenomenological derivation of spectral scaling laws, summarised in section \ref{summary-cascades}, we have given derivations of exact third-order laws for the forward-cascade range and for the inverse-cascade range (equations \ref{four-fifths-iso} and \ref{four-fifths-E-iso}).  These derivations are based upon general considerations of symmetries of the gyrokinetic equation, and the outcome is valid uniformly at all scales, above and below the Larmor radius.

Combining the exact third-order results with the dual self-similarity hypothesis (equation \ref{dual-hypothesis}), we have been able to reproduce the phenomenological scaling laws for both the forward-cascade range and the inverse-cascade range (section \ref{spectral-theory-sec}).  The dual self-similarity hypothesis is also used to predict scaling laws for the phase-space spectrum of free energy.  To describe the velocity-space structure, we have introduced a spectral formalism based on the zeroth-order Hankel transformation (section \ref{formalism-sec}).  This novel spectral treatment of perpendicular velocity space may in general prove useful for theoretical and numerical applications in gyrokinetics.  The works \cite{tatsuno-prl, tatsuno-pop} give strong numerical evidence of the $k$-space scalings, first predicted by \cite{schek-ppcf, schekochihin}, and our new predictions for the phase-space Fourier-Hankel spectrum $W_g(k, p)$ in the forward-cascade range.  Numerical investigation of other novel predictions of this paper, including the exact laws and the spectral scalings for the inverse cascade, is left for the future.

\subsection{Forcing and universality}

This paper explores the fundamental nonlinear processes that govern the turbulent evolution of a magnetised plasma.  For the sake of both simplicity and generality, we do not fix the particular mechanism that drives this turbulence.  It may, for instance, be driven by linear instabilities that are induced by large-scale (equilibrium-scale) gradients \cite[\eg, see][]{krommes}.  In this case, additional (linear) terms in the gyrokinetic equation must be included to account for the variation in the background equilibrium; simple phenomenological arguments indicate that the additional terms do not invalidate the inertial range assumption at fine scales.  The details of these arguments are left for future work.  

We think that the sub-Larmor scalings derived in this paper are likely to turn out to be universal properties of gyrokinetic turbulence.  We are encouraged in this view by the spectra of potential fluctuations rather similar to ours having been reported by \cite{gorler-published} in their three-dimensional simulations of tokamak turbulence.

\subsection{Fusion}\label{fusion-concl-sec}

It is often noted that the macroscopic properties of a fusion plasma (\ie transport of heat and particles) are most directly influenced by scales of the energy injection or larger (the Larmor scale in fusion plasmas).  This has meant that inertial-range (sub-Larmor-scale) theory is not a focus of present-day magnetically-confined fusion research.  Yet, understanding the phase-space cascade is central to understanding the kinetic nature of nonlinear interactions.  Furthermore, it is a robust mechanism for dissipation in gyrokinetics.  Indeed, the limit of weak collisionality will be increasingly relevant in the scaling to larger machines.  In this limit, the dissipation scale (equation \ref{dissip-scale-eqn}) decreases, while the nonlinear phase-space cascade ensures that fluctuations access this scale \cite[in the limit of small collisionality, the (free energy) dissipation rate approaches a non-zero constant, as demonstrated numerically by][]{tatsuno-prl}.  The ``practical'' application of inertial-range studies may, like in fluid turbulence, be in the development of subgrid-scale models for large-eddy simulation and the testing of nonlinear simulation codes.

\subsection{Future work}\label{future-work-sec}

There are some issues not addressed by this work that should be given high priority.  A more complete investigation of the gyrokinetic turbulence cascade would include more physical content, such as electromagnetic affects and the correct linear instability drives alluded to above.  Electromagnetism introduces a fluctuating magnetic field into the nonlinearity, which couples to the distribution function through Amp\`{e}re's law and will enrich the problem substantially, \eg, populating the sub-Larmor range with kinetic Alfv\'{e}n waves \cite[][]{schekochihin}.

Effects of the third spatial dimension must be investigated.  For instance, the assumption that the parallel streaming term (see equation \ref{gyro_full}) may be neglected requires careful justification.  In the context of tokamak turbulence, a conventional assumption is that one may take the size of $k_{\parallel}$ to correspond to the distance along a field line from the stable side to the unstable side of the toroidal magnetic surface.  If this assumption holds, the nonlinearity will dominate over the parallel streaming term for sufficiently small (perpendicular) scales.  In a more general context (\ie not specific to fusion), the term is believed to play a special role in the Alfv\'{e}n wave cascade --- which can be treated within gyrokinetic theory.  In this case, it is believed that a critical balance condition is satisfied where the generation of finer scales in $k_{\parallel}$ ensures that the parallel term must be included (at dominant order) at all scales in the cascade forward \cite[][]{schekochihin}.  

The linear phase-mixing process, due to the parallel streaming term, generates fine scales in the parallel-velocity ($v_{\parallel}$) dependence of the distribution function.  The nonlinear simulations and theoretical work of \cite{watanabe} demonstrate how this parallel phase-mixing, taken in isolation from nonlinear phase mixing, plays a role in the turbulent steady state.  However, the question of how exactly the two phase mixing processes coexist is open.  It has been heuristically argued by \cite{schek-ppcf} that the linear phase-mixing process is much weaker than nonlinear phase-mixing at small scales.  If that is true, the scaling theory developed in this paper should still hold for the inertial range of the forward cascade in three-dimensional gyrokinetics.

In general, the present capabilities of numerical simulations allow a fully resolved investigation of (perpendicular) nonlinear phase-mixing or (parallel) linear phase-mixing, but not both.  Thus, there is currently an opportunity for theoretical investigation of the full five-dimensional theory (that is, three position dimensions and two velocity dimensions) to pave the way to the full understanding of the turbulent steady state in magnetised plasma turbulence.

\section*{Acknowledgements}

GGP acknowledges scholarship support from the Wolfgang Pauli Institute, Vienna.  GGP and TT were supported by the US DOE Center for Multiscale Plasma Dynamics, Grant No. DE-FC02-04ER54785; they also thank the Leverhulme Trust International Network for Magnetised Plasma Turbulence for travel support.  AAS was supported by an STFC Advanced Fellowship and by the STFC Grant ST/F002505/2.  We thank Bill Dorland, Greg Hammett, Jian-Zhou Zhu and three referees for helpful criticisms.

\appendix
\section{Normalisation}\label{norm-sec}

The gyrokinetic equation, equation \ref{gyro_full}, is cast in dimensionless variables.  The normalisation is summarised in the following table:

\vspace{.25cm}
\begin{center}
\begin{tabular}{llll}
$t \vth/L \rightarrow t,\;\;$ & $x/\rho \rightarrow x,\;\;$ & $y/\rho \rightarrow y,\;\;$ & $z/L \rightarrow z,\;\;$  \\ 
$v/\vth \rightarrow v,\;\;$ & $\phi q L/(T_0 \rho) \rightarrow \phi,\;\;$ & $h \vth^3 L/(n_0 \rho) \rightarrow h,\;\;$  & $F_0\vth^3/n_0 \rightarrow F_0.$
\end{tabular}
\end{center}
\vspace{.25cm}

\noindent We have adopted the following definitions: the equilibrium density and temperature of the species of interest are $n_0$ and $T_0$; the mass and charge are $m$ and $q$; the thermal velocity is $\vth = \sqrt{T_0/m_s}$; the Larmor radius is $\rho = \vth/\Omega$ where the Larmor (cyclotron) frequency is $\Omega = qB_0/m$.  $L$ is the reference scale length (\ie system size), satisfying $\rho/L \sim \epsilon \ll 1$ for consistency with gyrokinetic ordering --- no further mention of $L$ will be needed. 

\section{Response models for the non-kinetic species}\label{boltz-sec}

For the sake of generality, we treat both ion and electron (Larmor radius) scale turbulence, but with only one kinetic species.  For ion-scale turbulence, we solve the gyrokinetic equation for ions and assume that the perturbed electron distribution function is given by the electron Boltzmann response (normalised to ion units):

\begin{equation}
\delta f_e \approx -\frac{eT_{0i}}{qT_{0e}}\phi F_0,
\end{equation}

\noindent where $e$ is the magnitude of the electron charge, and $q$ is the ion charge.  As an alternative to the Boltzmann response, we also consider the `no response' model, $\delta f_e = 0$.  This is the exact two-dimensional solution, whereas the Boltzmann electron response is considered ``quasi two-dimensional'' because a small but non-vanishing parallel wavenumber is assumed.  A discussion of this can be found in \cite{hasegawa}.  Ion gyrokinetics with Boltzmann or no-response electrons is appropriate at all scales larger than the electron Larmor radius.

For the case of electron-scale turbulence (at the electron-Larmor-radius scale and smaller), the appropriate ion response model is again the Boltzmann response (normalised to electron units):

\begin{equation}
\delta f_i \approx -\frac{qT_{0e}}{eT_{0i}}\phi F_0.
\end{equation}

The choice of response model determines the parameter $\tau$ in the quasi-neutrality constraint, (equation \ref{qn-g}):

\begin{equation}
\label{tau-def}
\tau =
\begin{cases}
\frac{eT_{0i}}{qT_{0e}} & \text{\em kinetic ions, Boltzmann electrons,}\\
0 & \text{\em kinetic ions, no-response electrons,}\\
\frac{qT_{0e}}{eT_{0i}} & \text{\em kinetic electrons, Boltzmann ions.}
\end{cases}
\end{equation}

Note that although it may be useful to think of the ion-scale turbulence (\eg, for comparison with the Hasegawa--Mima turbulence), our results are more robust for electron-scale turbulence, as the ion Boltzmann response becomes more exact at high wavenumbers whereas the Boltzmann-electron approximation will be broken for high wavenumbers, approaching $k\rho_e \sim 1$.  

\section{Derivation of phase-space spectra}\label{spectra-deriv-sec}

\subsection{Derivation of equation \ref{skp-limit-scaling}}

In the limit $k \gg p$, the scaling of the spectrum $W_g(k,p)$ can be inferred from the scaling of $\Stwo$ (see equation \ref{sg-def-approx}) in the limit that $\ell_{} \ll \ell_v$.  Combining the scaling hypothesis given by equation \ref{dual-hypothesis} with the solution for the scaling index, $h_g = 1/6$ (equation \ref{hg-solution}), we may express $\Stwo$ as

\begin{equation}
\Stwo = |\ell_v|^{1/3}\; \psi\left(\frac{\ell}{|\ell_v|}\right),
\end{equation}

\noindent where $\psi$ is an unknown scaling function.  Now, assuming that for $\ell = 0$ and $\ell_v \neq 0$ the structure function is non-zero, $\Stwo(\ell = 0) \neq 0$, we have $\psi(0) = \psi_0 \neq 0$.  Taylor-expanding $\psi$ around $\ell_{}/\ell_v = 0$ we obtain

\begin{equation}
\Stwo \approx |\ell_v|^{1/3}\left[\psi_0 + \frac{\ell_{}}{|\ell_v|}\psi_0^{\prime}\right].
\label{s2-totic}
\end{equation}

\noindent When we substitute this into the expression for $W_{g1}(k,p)$, equation \ref{sg-def-approx}, the leading order term is negligible for $k \neq 0$.  Therefore,

\begin{align}
W_{g1}(k,p) &\approx -\frac{k\psi_0^{\prime}}{8}  \int \ell_{} d\ell_{}\; \bar{v} d\bar{v} \; d\ell_v J_0(k\ell_{})\cos(p\ell_v)\;\frac{\ell_{}}{|\ell_v|^{2/3}}.
\label{sg-pllk-scaling}
\end{align}

\noindent Equation \ref{sg-pllk-scaling} implies the power law $W_g(k,p) \propto p^{-1/3}k^{-2}$.  Repeating this analysis for the limit $k \ll p$ completes the derivation of equation \ref{skp-limit-scaling}.

\subsection{Derivation of equations \ref{sg-k-scaling} and \ref{sg-p-scaling}}

Integrating equation \ref{sg-def} over $p$, we obtain (for non-zero $k$)

\begin{equation}
W_{g1}(k) \equiv \int dp W_{g1}(k,p) = -\pi \frac{k}{2} \int \ell_{}d\ell_{} v dv J_0(k\ell_{})\Stwo(\ell_{}, v, v).
\end{equation}

\noindent Integrating equation \ref{sg-def-approx} over $k$, we obtain

\begin{equation}
W_{g1}(p) \equiv \int dk W_{g1}(k,p) \approx -\frac{1}{8}\int \bar{v} d\bar{v} d\ell_{v}\cos(p\ell_{v})\Stwo(\ell = 0, \ell_v, \bar{v}).
\end{equation}

\noindent The scaling of $\Stwo$ for zero gyro-centre increment $\ell_{} = 0$ can be obtained from equation \ref{s2-totic}.  The scaling of $\Stwo$ for zero velocity increment $\ell_v$ is obtained analogously.  These expressions imply equations \ref{sg-k-scaling} and \ref{sg-p-scaling}.

\bibliographystyle{jfm}
\bibliography{2d-gk-bib-nn}

\end{document}